\documentclass[10pt,aps,twocolumn,preprintnumbers,prd,noshowpacs,nofootinbib,noshowkeys,floatfix,superscriptaddress]{revtex4-2}
\usepackage[normalem]{ulem}
\usepackage{amsmath}
\usepackage{ulem}
\newcommand\redout{\bgroup\markoverwith
{\textcolor{red}{\rule[.5ex]{5pt}{0.7pt}}}\ULon}
\usepackage{xcolor}
\usepackage{placeins}
\usepackage{comment}
\usepackage{graphicx}
\usepackage[colorlinks = true, 
            linktoc    = all, 
            linkcolor  = black, 
            citecolor  = blue, 
            urlcolor   = magenta]{hyperref} 

\begin{document}
\title{\texttt{BHAC-QGP}: three-dimensional MHD simulations of relativistic heavy-ion collisions \\ I. Methods and tests}
\date{\today}
\author{Markus Mayer}
\email[E-mail: ]{mmayer@itp.uni-frankfurt.de}
\affiliation{Goethe University Frankfurt, Max-von-Laue-Straße 1, 60438 Frankfurt am Main, Germany}
\author{Ashutosh Dash}
\affiliation{Goethe University Frankfurt, Max-von-Laue-Straße 1, 60438 Frankfurt am Main, Germany}
\author{Gabriele Inghirami}
\affiliation{GSI Helmholtzzentrum für Schwerionenforschung, Darmstadt, Germany}
\author{Hannah Elfner}
\affiliation{GSI Helmholtzzentrum für Schwerionenforschung, Darmstadt, Germany}
\affiliation{Goethe University Frankfurt, Max-von-Laue-Straße 1, 60438 Frankfurt am Main, Germany}
\affiliation{Frankfurt Institute for Advanced Studies, Ruth-Moufang-Straße 1, 60438 Frankfurt am Main, Germany}
\affiliation{Helmholtz Research Academy Hesse for FAIR, Max-von-Laue-Straße 12, 60438 Frankfurt am Main, Germany}
\author{Luciano Rezzolla}
\affiliation{Goethe University Frankfurt, Max-von-Laue-Straße 1, 60438 Frankfurt am Main, Germany}
\affiliation{Frankfurt Institute for Advanced Studies, Ruth-Moufang-Straße 1, 60438 Frankfurt am Main, Germany}
\affiliation{School of Mathematics, Trinity College, Dublin 2, Ireland}
\author{Dirk H.\ Rischke}
\affiliation{Goethe University Frankfurt, Max-von-Laue-Straße 1, 60438 Frankfurt am Main, Germany}
\affiliation{Helmholtz Research Academy Hesse for FAIR, Max-von-Laue-Straße 12, 60438 Frankfurt am Main, Germany}

\begin{abstract}
We present \texttt{BHAC-QGP}, a new numerical code to simulate the evolution of matter created in heavy-ion collisions in the presence of electromagnetic fields. 
It is derived from the Black Hole Accretion Code (\texttt{BHAC}), which has been designed to model astrophysical processes in a general-relativistic magnetohydrodynamical description. 
As the original Black Hole Accretion Code, \texttt{BHAC-QGP} benefits from the use of Adaptive Mesh Refinement (AMR), which allows us to dynamically adjust the resolution where necessary, and makes use of time-dependent Milne coordinates and the ultrarelativistic equation of state, $P = e/3$. 
We demonstrate that \texttt{BHAC-QGP} accurately passes a number of systematic and rigorous tests.
\end{abstract}
\maketitle
\section{Background}
The collision of two heavy ions at almost the speed of light produces an extremely hot and dense system that is commonly known as quark-gluon plasma (QGP). 
Quarks are approximately massless and no longer confined in this system. 
Heavy-ion collision experiments at the Relativistic Heavy-Ion Collider (RHIC) and at the Large Hadron Collider (LHC) produce such a system with the aim to clarify the structure of the phase diagram of strong-interaction matter. 
These experiments lead to the conclusion that the QGP behaves like an almost perfect fluid 
\cite{STAR:2005gfr, Jacobs:2004qv, PHENIX:2004vcz, PHOBOS:2004zne, Romatschke:2009im, 
Romatschke:2007mq, Heinz:2005bw, Heinz:2004ar, Baier:2006um, Song:2010mg, BRAHMS:2004adc, 
Dusling:2007gi, Luzum:2008cw}.
\par
A quite successful framework to model and describe the evolution of the produced matter is relativistic hydrodynamics \cite{Romatschke:2009im, Jaiswal:2016hex, Csernai:1982zz, Rischke:1995ir, Kasza:2018qah, Florkowski:2012mz, Florkowski:2010zz}. 
Indeed, second-order dissipative relativistic hydrodynamics is able to reproduce transverse-momentum spectra and especially the so-called elliptic flow, a quantity that describes the azimuthal anisotropy in momentum space of particles produced in non-central collisions \cite{Dumitru:1998es, Molnar:2004yh, Petersen:2015rra, Petersen:2014yqa, Heinz:2013th}. 
These measurements suggest that particles interact strongly and the fireball thermalizes rapidly.
\par
The protons that do not directly participate in the heavy-ion collision, the so-called spectators, generate strong electromagnetic fields \cite{Skokov:2009qp, Tuchin:2014hza, Tuchin:2013apa, Tuchin:2013ie, Bzdak:2011yy, Deng:2012pc, McLerran:2013hla, Voronyuk:2011jd, Dash:2023kvr}. 
Assuming to be in vacuum, the strength of these fields can be estimated with the Li\'{e}nard-Wiechert potentials \cite{Bzdak:2011yy, Tuchin:2013ie, Deng:2012pc} to reach magnetic field strengths of order $B \sim 10^{18}~\mathrm{G}$ in a typical gold-gold collision at RHIC energies.  
Thus, ultrarelativistic heavy-ion collisions create the strongest magnetic fields in the Universe. 
Such strong electromagnetic fields may have a significant impact on the dynamics of heavy-ion collisions as they will induce a charge current in the QGP. 
Furthermore, chiral symmetry is restored in the QGP, because the light quark masses are much smaller than any other energy scale in the problem, and therefore they can be considered to have a well-defined chirality and, thus, also helicity. 
Due to fluctuations, it may be possible that the number of left-handed particles is not equal to the number of right-handed particles in the QGP. 
For a system with such a chiral imbalance, sufficiently strong electromagnetic fields can lead to novel effects, the ``chiral magnetic effect'' (CME) being the most prominent one \cite{Kharzeev:2007jp, Kharzeev:2015znc, Kharzeev:2010gd, 
Fukushima:2008xe, Deng:2012pc, Roy:2015kma, Tuchin:2014iua, Stephanov:2012ki, Kharzeev:2012ph}.
\par
Additionally, a non-central heavy-ion collision can create a QGP with extremely large angular momentum, leading to the most vortical fluid ever observed, with vorticities of the order of $\sim 10^{-22} ~\text{s}^{-1}$ \cite{Becattini:2020ngo, Becattini:2015ska, STAR:2017ckg, McInnes:2017rxu, Csernai:2014hva, Csernai:2013bqa, Liang:2022ekv, Becattini:2007sr}. 
Due to spin-orbit coupling, this vorticity can be converted into a polarization of particles \cite{Gao:2020lxh, Weickgenannt:2020aaf, Liang:2004ph, Betz:2007kg, STAR:2017ckg, STAR:2018gyt, ALICE:2019aid, Becattini:2013vja} and give rise to the so-called ``chiral vortical effect'' (CVE) \cite{Kharzeev:2007jp, Kharzeev:2015znc, Roy:2015kma, Becattini:2020ngo, Stephanov:2012ki}. 
\par
Because of these novel effects, it is important to understand the impact of electromagnetic fields on the dynamics of the system. 
This problem needs to be addressed by numerical codes solving the (3+1)-dimensional equations of relativistic magnetohydrodynamics (RMHD) \cite{Porth:2016rfi, DelZanna:2007pk, DelZanna:2013eua, Rolando:2014vea, Karpenko:2013wva, Mignone:2011fd}. 
In order to investigate the evolution of the QGP in the presence of external magnetic fields, we have developed \texttt{BHAC-QGP}, as a suitable adaption of the Black Hole Accretion Code (\texttt{BHAC}), which was originally developed for astrophysical applications, solving the equations of RMHD in 3+1 dimensions and with a second-order formulation of the equations of motion \cite{Porth:2016rfi}. 
\par
There exist a number of numerical hydrodynamics codes which are used to study heavy-ion collisions. 
All of these codes differ in their algorithms and techniques to simulate the collision process.
The \texttt{iEBE-VISHNU} code, for example, performs (2+1)-dimensional event-by-event simulations for relativistic heavy-ion collisions using viscous hydrodynamics \cite{Shen:2014vra}. 
\texttt{MUSIC} uses the Kurganov-Tadmor algorithm to study (3+1)-dimensional heavy-ion collisions using viscous hydrodynamics on an unstructured mesh \cite{Schenke:2010nt}. 
Similarly, \texttt{VHLLE} solves the equations of relativistic viscous hydrodynamics in the Israel-Stewart framework \cite{Karpenko:2013wva}.
\texttt{CLVisc} can be used to calculate the (3+1)-dimensional viscous hydrodynamic equations on Graphic Processing Units (GPUs) using the Open Computing Language (OpenCL) \cite{Pang:2018zzo}.
Okamoto et al.\ developed a Godunov-type relativistic-hydrodynamics code that is optimized specifically for Milne coordinates \cite{Okamoto:2016pbc}.
\texttt{PLUTO} is able to solve the (3+1)-dimensional equations of magnetohydrodynamics (MHD). 
It benefits from Adaptive Mesh Refinement (AMR), but is mainly designed to study astrophysical processes in flat spacetime \cite{Mignone:2007iw, Mignone:2011fd}.
With \texttt{ECHO-QGP}, on the other hand, it is possible to study heavy-ion collisions under the influence of magnetic fields by solving the (3+1)-dimensional equations of ideal RMHD \cite{DelZanna:2013eua, Inghirami:2016iru}.
With a recent improvement of the code, heavy-ion collisions can even be modeled in the regime of resistive RMHD \cite{Nakamura:2022wqr}. 
With \texttt{BHAC-QGP}, we aim to develop a code for heavy-ion collisions that solves the (3+1)-dimensional RMHD equations while using AMR.
At first, we will restrict ourselves to ideal RMHD, but ultimately we want to extend this towards second-order dissipative RMHD, for a first attempt in this direction see Ref.\ \cite{Dash:2022xkz}.
\par
This paper is intended to explain the formalism underlying \texttt{BHAC-QGP}, as well as the extensions as compared to the orginal \texttt{BHAC} code. 
Specifically, in Sec.\ \ref{Formalism} we discuss the framework of ideal RMHD as well as Milne coordinates. 
These are very well suited to describe the evolution of a QGP. 
In this section we will also describe the numerical formalism and the numerical methods applied in \texttt{BHAC-QGP}. 
In Sec.\ \ref{NumericalTests} we present and discuss a series of tests that demonstrate the viability and capabilities of \texttt{BHAC-QGP} and we conclude in Sec.\ \ref{ConclusionOutlook} with a summary and an outlook.
\par
Throughout this article we use natural Heaviside-Lorentz units, $\hbar = c = k_B = \epsilon_0 = \mu_0 = 1$. 
In these units the electric charge is dimensionless $e := \sqrt{4 \pi \alpha \hbar c} \approx 0.303$, where $\alpha \approx 1/137$ is the fine-structure constant. 
As signature of the metric tensor, we choose $\left(-,+,+,+\right)$. 
We use Greek indices to indicate the components of a four-vector, while Latin indices range from 1 to 3 for its spatial components. 
Bold letters indicate three-vectors. 
The scalar and vector products between three-vectors are as usual denoted by $\cdot$ and $\times$, respectively.
Furthermore, we will use the Levi-Civita tensor with $\epsilon^{0123} = 1$. 
\section{Formalism}
\label{Formalism}
  \subsection{Ideal relativistic magnetohydrodynamics}
RMHD is a successful framework to study the dynamics of electrically conducting fluids by 
coupling hydrodynamics with Maxwell's equations. The idea of MHD is that the magnetic fields 
induce currents in the moving conductive fluid, which then create forces on the fluid, which, 
in turn, can produce changes in the magnetic field 
\cite{Anile:1989rel, davidson:2016imd, Font:2007zza, Spruit:2013ud}. 
\par
A (relativistic) fluid can be described in terms of charge four-currents $N^{\mu}_i$, where $i = 1, 2, \ldots $ labels the currents of the associated conserved charges, and an energy-momentum tensor $T^{\mu \nu}$, such that the basic equations of hydrodynamics are given by the conservation of charges as well as total energy and momentum (see, e.g., Refs.\ \cite{rezzolla:2013rhd, csernai1994:ihc})
  \begin{eqnarray}
    \nabla_{\mu} N_i^{\mu} &=& 0    \;, \label{RMHD1}  \\[4pt]
    \nabla_{\mu} T^{\mu \nu} &=& 0  \;, \label{RMHD2}
  \end{eqnarray}
where $\nabla_{\mu}$ represents the covariant derivative.  
In heavy-ion collisions, for example, not only the electric charge, but also the baryon number and the strangeness are conserved. 
The total energy-momentum tensor can be decomposed into a matter part $T_m^{\mu \nu}$ and a field part $T_f^{\mu \nu}$,
  \begin{equation}
    T^{\mu \nu} = T_m^{\mu \nu} + T_f^{\mu \nu}  \;. \label{RMHD3}
  \end{equation}
In the case of an ideal fluid the conserved four-currents and the energy-momentum tensor can be decomposed as (see, e.g., Ref.\ \cite{rezzolla:2013rhd})
  \begin{eqnarray}
    N_i^{\mu} &:=& n_i u^{\mu}  \;, \label{RMHD4}  \\[4pt]
    T_m^{\mu \nu} &:=& e u^{\mu} u^{\nu} + P \Delta^{\mu \nu}  \;, \label{RMHD5}
  \end{eqnarray}
where $u^\mu$ is the four-velocity of the fluid, $n_i := -N_i^{\mu} u_{\mu}$ is the charge density associated with the charge current $N_i^\mu$ in the local rest frame (LRF) of the fluid, and $e := T_m^{\mu \nu} u_{\mu} u_{\nu}$ is the energy density in that frame. 
Furthermore, $P := T_m^{\mu \nu} \Delta_{\mu \nu}/3$ represents the pressure of the fluid, where $\Delta_{\mu \nu} := g_{\mu \nu} + u_{\mu} u_{\nu}$ is the projection operator onto the three-space orthogonal to the fluid velocity $u^\mu$. 
The fluid four-velocity can be written as $u^{\mu} = \left(\gamma, \gamma \boldsymbol{v}\right)^T$, where $\boldsymbol{v}$ is the three-velocity. 
It is a time-like, normalized vector, $u_{\mu} u^{\mu} = -1$, with $\gamma = \left(1-\boldsymbol{v}^{\,2}\right){}^{-1/2}$ being the Lorentz factor. 
\par
On the other hand, the electromagnetic field obeys Maxwell's equations
  \begin{eqnarray}
    \nabla_{\mu} F^{\mu \nu} &=& -J^{\nu}  \;, \label{RMHD6}   \\[4pt]
    \nabla_{\mu} {}^{*}F^{\mu \nu} &=& 0   \;, \label{RMHD7}
  \end{eqnarray}
where $F^{\mu \nu}$ is the antisymmetric electromagnetic Faraday tensor, ${}^{*}F^{\mu \nu} = \epsilon^{\mu \nu \alpha \beta} F_{\alpha \beta}/2$ is its dual, and $J^{\mu}$ characterizes the total electric charge four-current. 
Similar to the energy-momentum tensor, the Faraday tensor and its dual can be decomposed with respect to $u^\mu$ as
  \begin{eqnarray}
    F^{\mu \nu} &:=& u^{\mu} e^{\nu} - u^{\nu} e^{\mu} + \epsilon^{\mu \nu \alpha \beta} u_{\alpha} b_{\beta}  \;,   \label{RMHD8} \\[4pt]
    {}^{*}F^{\mu \nu} &:=& u^{\mu} b^{\nu} - u^{\nu} b^{\mu} - \epsilon^{\mu \nu \alpha \beta} u_{\alpha} e_{\beta}  \;,  \label{RMHD9}
  \end{eqnarray}
where $e^{\mu}$ and $b^{\mu}$ are the electric field and the magnetic field in the comoving frame, respectively, 
  \begin{eqnarray}
    e^{\mu} &:=& F^{\mu \nu} u_{\nu} = \left(\gamma \, \boldsymbol{v} \cdot \boldsymbol{E}, \gamma \, \boldsymbol{E} + \gamma \, \boldsymbol{v} \times \boldsymbol{B}\right)^T   \;, \label{RMHD10}  \\[4pt]
    b^{\mu} &:=& {}^{*}F^{\mu \nu} u_{\nu} = \left(\gamma \, \boldsymbol{v} \cdot \boldsymbol{B}, \gamma \, \boldsymbol{B} - \gamma \, \boldsymbol{v} \times \boldsymbol{E}\right)^T \;. \label{RMHD11} 
  \end{eqnarray}
Here, $\boldsymbol{E}$ and $\boldsymbol{B}$ indicate the spatial electric and magnetic fields in the lab frame. 
Equations (\ref{RMHD10}) and (\ref{RMHD11}) imply that $e^{\mu}$ and $b^{\mu}$ are space-like vectors, $e^{\mu} e_{\mu} > 0, b^{\mu} b_{\mu} > 0$.
\par
The coupling of the electromagnetic field to the QGP is ensured by an appropriately adapted Ohm's law, which relates the electric-charge current to the electromagnetic field. 
In general, there are various contributions to the current, but hereafter only the conduction current $j^{\mu}_{\text{cond}}$ and the induction current $j^{\mu}_{\text{ind}}$ are taken into account. 
In this case, the total charge four-current takes the form
  \begin{equation}
    J^{\mu} = j^{\mu}_{\text{cond}} + j^{\mu}_{\text{ind}} = \rho_{\text{q}} u^{\mu} + \sigma_{E}^{\mu \nu} e_{\nu}  \;,  \label{RMHD12}
  \end{equation}
where $\rho_{\text{q}}$ represents the electric-charge density in the comoving frame. 
In the case of $\rho_{\text{q}} = 0$ Eq.\ (\ref{RMHD12}) reduces to the ideal Ohm's law. 
Moreover, $\sigma^{\mu \nu}_E$ is the conductivity tensor of the QGP, which can be replaced by a scalar function in the case of an isotropic conductivity, $\sigma_{E}^{\mu \nu} e_{\nu} \rightarrow  \sigma_{E}  e^{\mu}$, where $ \sigma_{E}$ is the electric conductivity of the QGP. 
Hence, for an electrically neutral QGP with isotropic conductivity the total charge four-current takes the form
  \begin{equation}
    J^{\mu} =  \sigma_{E} e^{\mu}  \;. \label{RMHD13}
  \end{equation}
\par
The equations of RMHD simplify considerably when the electric conductivity is assumed to be infinite, $\sigma_E \rightarrow \infty$, which is equivalent to a vanishing resistivity, $\eta_E := 1/\sigma_E  \rightarrow 0$, and is referred to as the ideal-MHD limit.
Because of the vanishing resistivity, the electric field must vanish in the comoving frame,
\begin{equation}
    e^{\mu} = 0  \;. \label{RMHD14}
\end{equation}
Therefore, the inhomogeneous Faraday law (\ref{RMHD6}) is not required and the electric fields are uniquely given as functions of the velocities and magnetic fields, 
  \begin{equation}
    \boldsymbol{E} = - \boldsymbol{v} \times \boldsymbol{B}  \;, \label{RMHD15}
  \end{equation}
which follows from the spatial part of Eq.\ (\ref{RMHD10}). 
With the help of relation (\ref{RMHD15}), the magnetic field four-vector (\ref{RMHD11}) can then be written as 
  \begin{equation}
    b^{\mu} = \left(\gamma \, \boldsymbol{v} \cdot \boldsymbol{B}, \dfrac{\boldsymbol{B}}{\gamma} + \gamma \, \boldsymbol{v} \left(\boldsymbol{v} \cdot \boldsymbol{B}\right)\right)^T  \;. \label{RMHD16}
  \end{equation}
The equation governing the evolution of the magnetic field (\ref{RMHD9}) in the ideal-MHD limit is then given by
  \begin{equation}
    \nabla_{\mu} \left(u^{\mu} b^{\nu} - u^{\nu} b^{\mu}\right) = 0  \;, \label{RMHD17} 
  \end{equation}
and the electromagnetic energy-momentum tensor reads as
  \begin{eqnarray}
    T_{f}^{\mu \nu} &=& F^{\mu \alpha} {F^{\nu}}_{\alpha} - \dfrac{1}{4} g^{\mu \nu} F^{\alpha \beta} F_{\alpha \beta}  \notag \\[4pt]
    &=& b^{2} u^{\mu} u^{\nu} + \dfrac{1}{2} b^{2} g^{\mu \nu} - b^{\mu} b^{\nu}  \;, \label{RMHD18}
  \end{eqnarray}
where $b^2 := b^{\mu} b_{\mu} = \boldsymbol{B}^{2}/\gamma^2 + \left(\boldsymbol{v} \cdot \boldsymbol{B}\right)^2$ is the magnetic field strength. 
The sum of the fluid's energy-momentum tensor (\ref{RMHD5}) and the electromagnetic energy-momentum tensor (\ref{RMHD18}) then yields the energy-momentum tensor of a perfect, magnetized relativistic fluid in the ideal-MHD limit
  \begin{equation}
    T^{\mu \nu} := \left(e + \dfrac{b^{2}}{2}\right) u^{\mu} u^{\nu} + \left(P + \dfrac{b^{2}}{2}\right) \Delta^{\mu \nu} - b^{\mu} b^{\nu}  \;. \label{RMHD19}
  \end{equation}
In this context, it is natural to introduce the total energy density and the total pressure
  \begin{equation}
    e_{\text{tot}} := e + \dfrac{b^2}{2}
   \;, \qquad
    P_{\text{tot}} := P + P_{\text{mag}}  \;, \label{RMHD20}
  \end{equation}
where $P_{\text{mag}} := b^2/2$ is the magnetic pressure. 
\par
Equations (\ref{RMHD1}), (\ref{RMHD2}), and (\ref{RMHD17}) form the system of ideal-RMHD equations with the unknowns $n, e, P, u^{\mu}$, and $b^{\mu}$. 
This system of equations is closed by an appropriate equation of state, which relates the thermodynamic quantities, usually $P = P\left(e,n\right)$. 
Furthermore, in the case of local thermodynamic equilibrium, the first law of thermodynamics
  \begin{equation}
    \text{d}e = T ~\text{d}s + \mu ~\text{d}n  \label{RMHD21}
  \end{equation}
can be used to formulate the entropy conservation equation as
  \begin{equation}
    u^{\mu} \partial_{\mu} s + s ~\partial_{\mu} u^{\mu} = 0  \;, \label{RMHD22}
  \end{equation}
with $s$ being the entropy density and $T, \mu$ being the temperature and chemical potential, respectively,
  \begin{equation}
    T := \left(\dfrac{\partial e}{\partial s}\right)_n  \;,
  \qquad 
    \mu := \left(\dfrac{\partial e}{\partial n}\right)_s  \;. \label{RMHD23}
  \end{equation}
  \subsection{Bjorken's model \& Milne coordinates}
Due to its complexity, the system of ideal-RMHD equations (\ref{RMHD1}), (\ref{RMHD2}), and (\ref{RMHD17}) cannot be solved analytically in general. 
However, an exception is offered by Bjorken's model \cite{Bjorken:1982qr}, which describes a rather idealized situation, but is very helpful in gaining a general understanding of the hydrodynamics of heavy-ion collisions. 
Bjorken's model considers a fluid which is expanding only in the beam direction, usually taken to be the $z$-axis, with velocity $v^z = z/t$, where $z$ and $t$ are spatial and temporal Cartesian coordinates. 
Thus, the four-velocity of this fluid is, per definition, given by
  \begin{equation}
    u^{\mu} := \left(\gamma, 0, 0, \gamma \, z/t\right)^T  \;. \label{MilneCoords1}
  \end{equation}
As a consequence, all fluid elements maintain their initial velocity without further acceleration. 
This assumption is equivalent to saying that the spacetime evolution of the system is the same in any reference frame, so that the expansion of this fluid is boost-invariant.
\par
Rather than using standard Cartesian coordinates $\left(t, x, y, z\right)$ for this boost-invariant flow, it is more convenient to use Milne coordinates $\left(\tau, x, y, \eta_{S}\right)$, also known as Bjorken coordinates, defined via the following transformation:
  \begin{eqnarray}
    t &= &\tau \cosh  \eta_{S}  \;, \label{MilneCoords2} \\[4pt]
    z &= &\tau \sinh \eta_{S}   \;, \label{MilneCoords3}
  \end{eqnarray}
where $\tau$ is the proper time and $\eta_{S}$ the spacetime rapidity, namely
  \begin{eqnarray}
    \tau   &:= &\sqrt{t^2 - z^2}  \;, \label{MilneCoords4} \\[4pt]
    \eta_{S} &:= &\frac{1}{2} \, \text{ln}\!\left(\frac{t + z}{t - z}\right)     \;. \label{MilneCoords5}
  \end{eqnarray}
The spacetime rapidity $\eta_S$ is only defined in the timelike region, where $t > \vert z \vert$. 
The resulting Minkowski metric in Milne coordinates is given by
  \begin{equation}
    \tilde{g}^{\mu \nu} = \text{diag}\!\left(-1, 1, 1, \dfrac{1}{\tau^2}\right)  \;. \label{MilneCoords6}
  \end{equation}
From now on, we will use a tilde to refer to quantities in Milne coordinates. 
The four-velocity (\ref{MilneCoords1}) simplifies in Milne coordinates to
  \begin{equation}
    \tilde{u}^{\mu} = \left(1, 0, 0, 0\right)^T  \;, \label{MilneCoords10}
  \end{equation}
which corresponds to the four-velocity of the comoving frame. 
Consequently, the so-called expansion scalar $\theta$ and the substantial time derivative $\mathrm{D}/\mathrm{D} t$ in Milne coordinates are written as follows:
  \begin{equation}
    \theta := \nabla^{\mu} \tilde{u}_{\mu} = \dfrac{1}{\tau} \;,
  \qquad
    \dfrac{\mathrm{D}}{\mathrm{D} t} := \tilde{u}_{\mu} \nabla^{\mu} = \dfrac{\partial}{\partial \tau}  \;.  \label{MilneCoords11}
  \end{equation}   
\par
By assuming a magnetic field that is transverse to the fluid expansion, i.e., $b^{\mu} = \left(0, b^x, b^y, 0\right)^T$, it is possible to deduce the following energy conservation equation in Milne coordinates:
  \begin{equation}
    \dfrac{\partial}{\partial \tau} \left(e + \dfrac{b^2}{2}\right) = -\dfrac{e + P + b^2}{\tau}  \;. \label{MilneCoords12}
  \end{equation}
Equation (\ref{MilneCoords12}), which is obtained by contracting the energy-momentum conservation (\ref{RMHD2}) with the fluid-velocity (\ref{MilneCoords10}), states that in ideal MHD the decrease of the total energy density with time is proportional to the sum of fluid energy density $e$, pressure $P$, and the magnetic field strength $b^2$, divided by the proper time $\tau$. 
The evolution equation for the total energy density is independent of the spacetime rapidity $\eta_S$, thus emphasizing the boost-invariance property. 
The solution of the energy conservation equation (\ref{MilneCoords12}) requires an equation of state which relates the pressure to the energy density and the charge density, and the knowledge of the evolution of the magnetic field. 
Maxwell's equations (\ref{RMHD17}) can be written in the lab frame as
  \begin{eqnarray}
    \boldsymbol{\nabla} \cdot \boldsymbol{B} &=& 0  \;, \label{MilneCoords13} \\[4pt]
    \dfrac{\text{D} \boldsymbol{B}}{\text{D} t} &=& \left(\boldsymbol{B} \cdot  \boldsymbol{\nabla}\right)  \boldsymbol{v} - \boldsymbol{B} \, \left(\boldsymbol{\nabla} \cdot \boldsymbol{v}\right)  \;, \label{MilneCoords14}
  \end{eqnarray}
where Eq.\ (\ref{MilneCoords13}) is the well-known divergence constraint, while Eq.\ (\ref{MilneCoords14}) is the so-called induction equation. 
With the help of the continuity equation (\ref{RMHD1}) for the density $n$,
  \begin{equation}
    \dfrac{\text{D} n}{\text{D} t} = -n \left(\boldsymbol{\nabla} \cdot \boldsymbol{v}\right)  \;, \label{MilneCoords15}
  \end{equation}
the induction equation (\ref{MilneCoords14}) can be rewritten as \cite{Roy:2015kma}
  \begin{equation}
    \dfrac{\text{D}}{\text{D} t} \left(\dfrac{\boldsymbol{B}}{n}\right) = \dfrac{\boldsymbol{B}}{n} \left(\boldsymbol{\nabla} \cdot \boldsymbol{v}\right)  \;, \label{MilneCoords16}
  \end{equation}
which is also known as the frozen-flux theorem, according to which the magnetic field lines are simply advected with the fluid. 
In Milne coordinates, where $\boldsymbol{v} = \boldsymbol{0}$, this frozen-flux theorem becomes
  \begin{equation}
    \dfrac{\partial}{\partial \tau} \left(\dfrac{\boldsymbol{B}}{n}\right) = 0  \;, \label{MilneCoords17}
  \end{equation}
whose solution is obviously $\boldsymbol{B}\!\left(\tau\right)/\boldsymbol{B}_0 = n\!\left(\tau\right)/n_0$, with $\boldsymbol{B}_0 := \boldsymbol{B}\!\left(\tau = \tau_0\right)$ and $n_0 := n\!\left(\tau = \tau_0\right)$. 
Additionally, the continuity equation (\ref{RMHD1}) for the density in Milne coordinates reads as
  \begin{equation}
    \dfrac{\partial n}{\partial \tau} = -\dfrac{n}{\tau}  \;. \label{MilneCoords18}
  \end{equation}
The solution of Eq.~(\ref{MilneCoords18}) is trivial, $n\left(\tau\right)/n_0 = \tau_0/\tau$, and this allows us to rewrite the solution of Eq.\ (\ref{MilneCoords17}) as
  \begin{equation}
    B\!\left(\tau\right) = B_0 \, \dfrac{\tau_0}{\tau}    \;. \label{MilneCoords19}
  \end{equation}
A suitable choice of the equation of state, $P = P(e,n)$, then allows to solve the energy conservation equation (\ref{MilneCoords12}).
  \subsection{\texttt{BHAC} \& \texttt{BHAC-QGP}}
The Black Hole Accretion Code (\texttt{BHAC}) solves the equations of ideal general-relativistic MHD (GRMHD), namely Eqs.\ (\ref{RMHD1}), (\ref{RMHD2}), and (\ref{RMHD17}), in arbitrary but time-independent spacetimes \cite{Porth:2016rfi, Olivares:2019dsc}. 
For this purpose, \texttt{BHAC} uses second-order finite-volume methods, for which an appropriate separation of time and spatial components is necessary. 
This can be achieved by ``foliating'' the spacetime in terms of non-intersecting spacelike hypersurfaces. 
On each hypersurface it is then possible to define a timelike four-vector normal to the hypersurface as
  \begin{equation}
    n^{\mu} := \dfrac{1}{\alpha} \, \left(1, -\beta^j\right)^T  \;,
  \qquad
    n_{\mu} := \left(-\alpha, 0_j\right)  \;. \label{BHAC1}
  \end{equation} 
Here, $\alpha$ is the so-called lapse function while the purely spatial vector $\beta^{\mu}$ is called shift vector. 
Since $n^{\mu} \, n_{\mu} = -1$, this four-vector defines also the frame of an Eulerian observer. 
An observer in the Eulerian frame is at rest, while an observer in the so-called Lagrangian frame is moving with the fluid. 
The lapse function $\alpha$ measures the rate of change of the coordinate time along the vector $n^{\mu}$. 
The introduction of such a normal vector $n^{\mu}$ allows the definition of the metric associated to each foliated hypersurface as
  \begin{equation}
    \gamma_{\mu \nu} := g_{\mu \nu} + n_{\mu} n_{\nu}  \;,
  \qquad
    \gamma^{\mu \nu} := g^{\mu \nu} + n^{\mu} n^{\nu}  \;. \label{BHAC2}
 \end{equation}
With the help of the lapse function $\alpha$ and the shift vector $\beta^{\mu}$ it is possible to express a generic line element $ds^2$ in 3+1 decomposition as \cite{Arnowitt:1962hi}
  \begin{eqnarray}
  \!\!\!\!\!\!
    ds^2 &=& \left(-\alpha^2 + \beta^i \beta_i\right) dt^2 + 2 \beta_i dx^i dt + \gamma_{ij} dx^i dx^j  \;. \label{BHAC3}
  \end{eqnarray}
This expression indicates that in the case of Minkowski and Milne coordinates $\alpha = 1$ and $\beta^{\mu} = 0$. 
Hence, the timelike four-vector defined in Eq.\ (\ref{BHAC1}) coincides with the four-velocity in the local rest frame and can be used as a temporal projection operator. 
Furthermore, the metric associated to each hypersurface coincides with the spatial projection tensor, i.e.,
  \begin{align}
    \Delta^{\mu \nu} =  g^{\mu \nu} + u^{\mu} u^{\nu}  \;. \label{BHAC4}
  \end{align}
Contracting the energy-momentum tensor (\ref{RMHD19}) twice with $n_{\mu} = (-1, 0_j)$ yields the total energy density as seen by the Eulerian observer
  \begin{equation}
    e^{\prime}_{\text{tot}} = w \gamma^2 - P + \dfrac{1}{2} \left[\boldsymbol{B}^2 \left(1 + \boldsymbol{v}^2\right) - \left(\boldsymbol{v} \cdot \boldsymbol{B}\right)^2\right]   \;, \label{BHAC5}
  \end{equation}
where we defined the relativistic enthalpy density as
  \begin{equation}
    w := e + P  \;, \label{BHAC6}
  \end{equation}
which is related to the specific enthalpy $h$ via $h := w/\rho$, where $\rho$ is the rest-mass density in the local rest frame. 
Instead, contracting the energy-momentum tensor (\ref{RMHD19}) with $n_{\mu}$ and $\gamma_{\nu \lambda}$ leads to the covariant three-momentum density of the Eulerian observer,
  \begin{equation}
    S_j = w \gamma^2 v_j + \boldsymbol{B}^2 v_j - B_j \left(\boldsymbol{v} \cdot \boldsymbol{B}\right)  \;. \label{BHAC7}
  \end{equation}
Similarly, contracting the energy-momentum tensor (\ref{RMHD19}) twice with the projection tensor $\gamma_{\mu \nu}$ yields the spatial part of the stress-energy tensor $W^{ij}$,
  \begin{align}
    W^{ij} &= S^i v^j + P_{\text{tot}} \Delta^{ij} - \dfrac{B^i B^j}{\gamma^2} - v^i B^j \left(\boldsymbol{v} \cdot \boldsymbol{B}\right)  \;. \label{BHAC8}
  \end{align}
After deriving these quantities, it is now possible to write the ideal RMHD equations in conservative form, that is
  \begin{equation}
    \partial_0 \left(\sqrt{g_s} \, \boldsymbol{U}\right) + \partial_i \left(\sqrt{g_s} \, \boldsymbol{F}^{i}\right) = \sqrt{g_s} \, \boldsymbol{S}  \;. \label{BHAC9}
  \end{equation}
Here, $g_s$ is the determinant of the spatial part of the metric $g^{\mu \nu}$, while $\boldsymbol{U}$ and $\boldsymbol{F}^{i}$ are conserved variables and fluxes, respectively. 
In \texttt{BHAC} and \texttt{BHAC-QGP}, these are defined as
  \begin{equation}
    \boldsymbol{U} = 
      \begin{pmatrix}
        D     \\[4pt]
        S_j   \\[4pt]
        \varepsilon \\[4pt]
        B^j
      \end{pmatrix}
 \;, \quad
    \boldsymbol{F}^{i} =
      \begin{pmatrix}
        v^i D       \\[4pt]
        S_j        \\[4pt]
        S^i - D v^i \\[4pt]
        v^i B^j - B^i v^j
      \end{pmatrix}  \;. \label{BHAC10}
  \end{equation}
where $D := \gamma \rho$ is the density in the Eulerian frame and $\varepsilon := e^{\prime}_{\text{tot}} - D$ is the rescaled energy density.
The choice to use $\varepsilon = e^{\prime}_{\text{tot}} - D$ in the energy conservation equation, instead of only $e^{\prime}_{\text{tot}}$, has a purely numerical motivation. 
The conservation of $\varepsilon$ as a combination of two conserved quantities is simply more accurate than that of $e^{\prime}_{\text{tot}}$ only, especially in regions of low energy density. 
Furthermore, $\varepsilon$ enables us to recover the Newtonian limit. 
Mathematically, the use of $\varepsilon$ is possible because a linear combination of two conserved variables is still a solution of the equations in conservative form. 
We note that instead of the rest-mass density $\rho$ any other density, for example the baryon number density $n_B$, can be used, i.e., $\rho \rightarrow n_B$ and $D \rightarrow \gamma n_B$. 
However, since the dimensions of $\rho$ and $n_B$ do not match, only $e^{\prime}_{\text{tot}}$ is then evolved in the energy conservation equation instead of $\varepsilon = e^{\prime}_{\text{tot}} - D$.
While $D$ is part of the conserved variables, $\rho$ belongs to the so-called primitive variables. 
How the conversion between these variables is done is described in more detail in Sec.\ \ref{sec:II.F}.
\par
Furthermore, $\boldsymbol{S}$ in Eq.\ (\ref{BHAC9}) represents the source terms. 
According to the so-called Valencia formulation, these are given as \cite{rezzolla:2013rhd}:
  \begin{equation}
    \boldsymbol{S} = 
      \begin{pmatrix}
        0 \\[4pt]
        \dfrac{1}{2}~\alpha W^{ik} \partial_j \gamma_{ik} + S_i \partial_j \beta^i - e^{\prime} \partial_j \alpha \\[4pt]
        \alpha W^{ij} K_{ij} - S^j \partial_j \alpha \\[8pt]
        0
      \end{pmatrix}  \;, \label{BHAC11}
  \end{equation}
where $K^{ij}$ is the extrinsic curvature. 
Basically, it measures how the three-dimensional hypersurface is curved with respect to four-dimensional spacetime. 
It is defined as \cite{rezzolla:2013rhd}
  \begin{equation}
    K_{\mu \nu} = -\nabla_{\mu} n_{\nu} - n_{\mu} n^{\alpha} \nabla_{\alpha} n_{\nu}  \;. \label{BHAC12}
  \end{equation}
In the case of Milne coordinates, the only surviving spatial component for the extrinsic curvature is $K_{33} = -\tau$, so that in Milne coordinates the source terms are
  \begin{equation}
    \tilde{\boldsymbol{S}} = 
      \begin{pmatrix}
        0 \\[4pt]
        0 \\[4pt]
        W^{33} \, K_{33} \\[8pt]
        0
      \end{pmatrix}  \;. \label{BHAC13}
  \end{equation}
The source term vanishes when ordinary Cartesian coordinates are used.
\par
Now, all the necessary quantities have been introduced to write down the evolution equations of \texttt{BHAC-QGP}. 
However, while for Cartesian coordinates $\sqrt{g_s} = 1$, for Milne coordinates $\sqrt{g_s} = \tau$, so that a time dependence is still hidden in the latter case.
But for these coordinates it is possible to write Eq.\ (\ref{BHAC9}) as
  \begin{eqnarray}
    \partial_{\tau} \left(\tau \, \boldsymbol{U}\right) + \partial_i \left({\tau} \, \boldsymbol{F}^{i}\right) &=& \tau \, \tilde{\boldsymbol{S}}  \;, \notag \\
    \Longleftrightarrow\quad \partial_{\tau} \boldsymbol{U} + \partial_i \boldsymbol{F}^{i} &=& \tilde{\boldsymbol{S}} - \dfrac{\boldsymbol{U}}{\tau}  \;. \label{BHAC14}
  \end{eqnarray}
Equation (\ref{BHAC14}) represents the evolution equations of \texttt{BHAC-QGP}. 
The term $\boldsymbol{U}/\tau$ on the right-hand side is an additional source term that is responsible for a decay of the conserved variables and is a consequence of the longitudinal expansion of the fluid.
  \subsection{Equation of state}
  \label{subsec:EoS}
In order to solve the set of ideal RMHD equations, an appropriate equation of state (EOS) is required. 
In the parent code \texttt{BHAC}, the EOS of an ideal gas and the EOS of a Synge gas are already implemented and described in Sec.\ 2.4 of Ref.\ \cite{Porth:2016rfi}. 
However, these equations of state are not suitable for describing the properties and behavior of matter in heavy-ion collisions.
In order to describe the QGP created in heavy-ion collisions, we neglect the (small) masses of the quarks and for the sake of simplicity assume classical statistics.
Thus, we implement an EOS for classical massless particles
  \begin{equation}
    P = \dfrac{e}{3} = \dfrac{g}{ \pi^2 \left(\hbar c\right)^3} ~T^4  \;,  \label{EOS1}
  \end{equation}
where the factor $g$ accounts for the number of internal degrees of freedom. 
In this paper we assume g = 37.
Although natural units are used, the factor $\left(\hbar c\right)^3 \approx \left(0.197 ~\mathrm{GeV} ~\mathrm{fm}\right)^3$ is carried along to be able to write the pressure in units of $\mathrm{GeV}/\mathrm{fm}^3$.
\par
The enthalpy density (\ref{BHAC6}) reads for a fluid of massless particles as
  \begin{equation}
    w = e + P = 4 P \;, \label{EOS2}
  \end{equation}
while the square of the speed of sound is
  \begin{align}
    c_s^2 = \dfrac{\partial P}{\partial e} = \dfrac{1}{3}  \;. \label{EOS3}
  \end{align}
Due to the vanishing mass, the rescaled energy density simplifies to $\varepsilon = e^{\prime}_{\text{tot}}$.
Therefore, in this case, the generic quantity $D$ can be replaced by any conserved quantity with the dimension of energy density.
For classical massless particles the entropy density follows from the first law of thermodynamics as $s = 4 \, n$, where $n$ is the particle density. 
Therefore, in this case it is also possible to evolve the particle density instead of the mass density $\rho$. 
The particle density for classical massless particles is
  \begin{equation}
    n = \dfrac{g}{\pi^2 \left(\hbar c\right)^3} ~T^3  \;. \label{EOS4}
  \end{equation}
  \subsection{Numerical methods}
As the original Black Hole Accretion Code, \texttt{BHAC-QGP} solves the evolution equations (\ref{BHAC6}) in a finite-volume formulation that is described in Chapter 2.2 of Ref.\ \cite{Porth:2016rfi}. 
Basically, the integral of the evolution equations (\ref{BHAC9}) over the spatial element of each cell $\int \mathrm{d}x^1 \, \mathrm{d}x^2 \, \mathrm{d}x^3$ is taken, which in the end leads to a semi-discrete equation for the average state in a cell. 
This finite-volume approach has second-order accuracy and is powered by the \texttt{MPI-AMRVAC} toolkit. 
This allows \texttt{BHAC} and thus also \texttt{BHAC-QGP} to use the computational infrastructure efficiently. 
\par
The integrations are performed numerically during the initialization phase using Simpson's fourth-order rule. 
In order to achieve a correspondingly high accuracy for the temporal updates as well, \texttt{BHAC-QGP} relies on the so-called ``method of lines'' (see, for instance, Ref.\ \cite{rezzolla:2013rhd}). 
Essentially, this leaves the problem continuous in time such that the set of partial differential equations becomes a set of ordinary differential equations. 
In this way, the method of lines acquires the accuracy order of the used time integrator. 
Many time-integration schemes are implemented in \texttt{BHAC} and \texttt{BHAC-QGP}, e.g., simple predictor-corrector, second-order Runge-Kutta RK2, third-order Runge-Kutta RK3, but also the strong stability-preserving $s$-step, $p$-order Runge-Kutta schemes SSPRK($s$,$p$), namely the schemes SSPRK(4,3) and SSPRK(5,4).
\par
A second-order finite-volume method requires numerical fluxes on the mid-point of the interface. 
These point-wise fluxes $\boldsymbol{F}^{i}$ are obtained by performing a limited reconstruction operation of the cell-averaged state to the interfaces at each substep. 
Stated differently, the fluid variables must be \textit{reconstructed} at the interface with the help of an appropriate spatial interpolation. 
\texttt{BHAC-QGP} takes over the reconstruction strategy from its parent code, which can be summarized as follows:
  \begin{enumerate}
    \item Compute the so-called primitive variables from the averages of the conserved variables located at the center of a cell.
    \item Use the reconstruction formula to obtain two representations for the state at the interface, one with left-based reconstruction stencil and one with right-based stencil.
    \item Convert the point-wise values back to their conserved states and use the resulting left and right states as input for the approximate Riemann solver.
  \end{enumerate}
There are various reconstruction schemes implemented in \texttt{BHAC} and \texttt{BHAC-QGP} \cite{Porth:2016rfi}: 
standard second-order total-variation diminishing (TVD) schemes like ``minmod'', ``vanLeer'', ``woodward'', or ``koren'', but also higher-order methods like ``PPM'' (third-order) and ``MP5'' (fifth-order). 
A higher accuracy of the spatial discretization usually improves the accuracy of the solution, as it reduces the diffusion of the scheme. 
However, the overall order of the finite-volume method will remain second order. 
\par
The reconstructed fluxes are then computed by solving approximate Riemann problems, which arise naturally at the interfaces. 
This strategy belongs to the widely used Godunov scheme. 
Implemented Riemann solvers are, for example, HLL (Harten–Lax–Van Leer) or TVDLF (Total variation-diminishing Lax-Friedrichs scheme). 
  \subsection{Primitive-variable recovery and entropy switch}
  \label{sec:II.F}
A problem of all RMHD codes is the non-linear inversion from conservative to primitive variables. 
The conserved variables of \texttt{BHAC-QGP} have been introduced in Eq.\ (\ref{BHAC10}),
  \begin{equation}
    \boldsymbol{U} = 
      \left(
        D,
        S_j,
        \varepsilon,
        B^j
      \right)^T  \;. \notag \label{con2prim1}
  \end{equation}
In order to calculate the fluxes $\boldsymbol{F}$ for these conserved variables along the interfaces of the computational cells, an additional set of so-called primitive variables is needed. 
These primitive variables are
  \begin{align}
    \boldsymbol{P}\left(\boldsymbol{U}\right) =
      \left(
        \rho,
        \gamma v_j,
        P,
        B^j
      \right)^T  \;. \label{con2prim2}
  \end{align}
The computation of the fluxes requires the velocity and the pressure from the primitive variables. 
But, \texttt{BHAC} and thus also \texttt{BHAC-QGP} do not store these primitive variables. 
Instead, the two auxiliary variables $\gamma$ and $\xi := \gamma^2 \rho h$ are stored in addition to the conserved variables $\boldsymbol{U}$. 
Knowledge of $\gamma$ and $\xi$ allows then a quick transformation to the primitive variables. 
In the case of an ultrarelativistic fluid, the auxiliary variable $\xi$ reads
  \begin{align}
    \xi = 4 P \gamma^2  \;. \label{con2prim3}
  \end{align} 
While the transformation $\boldsymbol{U}\left(\boldsymbol{P}\right)$ is trivial, the inversion $\boldsymbol{P}\left(\boldsymbol{U}\right)$ is much more complicated. 
No closed-form solution exists for the inversion from conserved to primitive variables, and hence the primitive variables must be found numerically. 
High Lorentz factors or strong magnetic fields may easily lead to errors in the values of the conserved variables. 
Consequently, a robust procedure is required. 
Two primary inversion strategies are implemented in \texttt{BHAC-QGP}, as in its parent code \cite{Porth:2016rfi, Ripperda:2019lsi}. 
\par
The inversion scheme labeled as ``2DW'' simultaneously solves Eq.\ (\ref{BHAC5}) and the square of the three-momentum (\ref{BHAC7}) \cite{Noble2006:prm, Qian:2016lyn, Bucciantini:2012sm}. 
To be more precise, \texttt{BHAC} and \texttt{BHAC-QGP} follow the strategy outlined in Refs.\ \cite{Noble2006:prm, DelZanna:2007pk}, where the equations are reduced to a set of two non-linear equations in the variables $\xi$ and $\chi := \boldsymbol{v}^2$. 
Multiplying the three-momentum (\ref{BHAC7}) with $B^j$ and calculating $\boldsymbol{S}^2$ yields 
  \begin{eqnarray}
     F_1\!\left(\chi, \xi\right) &:= &  \chi \left(\xi + \boldsymbol{B}^2\right)^2 - \dfrac{\left(\boldsymbol{S} \!\cdot\! \boldsymbol{B}\right)^2}{\xi^2} \left(2 \xi + \boldsymbol{B}^2\right) - \boldsymbol{S}^2  =0 \;, \notag \\ \label{con2prim8}
     \end{eqnarray}
as well as from Eqs.\ (\ref{BHAC5}) and (\ref{BHAC7})
\begin{eqnarray}
    F_2\!\left(\chi, \xi\right) &:= &  \xi - P + \dfrac{\boldsymbol{B}^2}{2} \left(1 + \chi\right) - \dfrac{\left(\boldsymbol{S} \!\cdot\! \boldsymbol{B}\right)^2}{2 \xi^2} - e^{\prime}_{\text{tot}} =0 \;. \notag \\
\label{con2prim9} 
  \end{eqnarray}
From these two equations, $\chi$ and $\xi$ can be found with a simultaneous two-dimensional Newton-Raphson root-finding procedure. 
\par
Instead of simultaneously finding the two roots via a two-dimensional root-finding scheme, it is also possible to derive $\chi = \chi\left(\xi\right)$ from Eq.\ (\ref{con2prim8}) and then find the root of the remaining equation $F_2\left[\chi\left(\xi\right), \xi\right] = 0$ with a one-dimensional Newton-Raphson procedure. 
However, with this routine the condition $\chi < 1$ must be ensured and it may happen that various iterations are required to solve the second equation. 
Once $\chi$ and $\xi$ are found, the required primitive variables are obtained from their relations to the conserved variables:
  \begin{eqnarray}
    \rho &=& D \, \left(1 - \chi\right)^{1/2}  \;,  \\
    v_j  &=& \dfrac{1}{\xi + \boldsymbol{B}^2} \, \left(S_j + B_j \, \dfrac{\boldsymbol{S} \cdot \boldsymbol{B}}{\xi}\right)  \;. \label{v_j}
  \end{eqnarray}
Since the pressure has to be specified by an EOS, in general no explicit expression can be given for it. 
Specifically, for a massless fluid, the relation reads according to Eq.\ (\ref{con2prim3}) as
  \begin{equation}
    P = \dfrac{\xi}{4} \, \left(1 - \chi\right)  \;.
  \end{equation}
\par
Another commonly used strategy, which is denoted as ``1DW'', is to solve the equation for the rescaled energy density (\ref{BHAC5}) \cite{Palenzuela:2008sf, Dionysopoulou:2012zv, vanderHolst:2008pg}, so that $\xi$ is a root of 
  \begin{equation}
    F\!\left(\xi\right) = \xi - e^{\prime}_{\text{tot}} - P + \dfrac{1}{2} \left[\boldsymbol{B}^2 \left(1 + \boldsymbol{v}^2\right) - \left(\boldsymbol{v} \cdot \boldsymbol{B}\right)^2\right] =0  \;.
  \end{equation}
Furthermore, an additional equation for the second auxiliary variable $\gamma$ is required. 
This relation can be extracted from Eq.\ (\ref{v_j}) as
  \begin{equation}
    \dfrac{1}{\gamma^2} = 1 - v^2 = 1 - \left(\dfrac{\boldsymbol{S} + \xi^{-1} \boldsymbol{B} \left(\boldsymbol{S} \cdot \boldsymbol{B}\right)}{\xi + \boldsymbol{B}^2}\right)^2  \;.
  \end{equation}
Once a new $\xi$ is found, the velocity can be computed from Eq.\ (\ref{BHAC7}) as
  \begin{equation}
    \boldsymbol{v} = \dfrac{\boldsymbol{S} + \xi^{-1} \boldsymbol{B} \left(\boldsymbol{S} \cdot \boldsymbol{B}\right)}{\xi + \boldsymbol{B}^2}  \;,
  \end{equation}
from which the density $\rho = D/\gamma$ can be obtained. 
Finally, the pressure follows from the respective EOS. 
\par
It is possible that both inversion strategies will fail in regions where the magnetic pressure is much stronger than the gas pressure, $P_{\text{mag}}/P \gtrsim 100$. 
Especially in heavy-ion collisions, the magnetic pressure outside the collision zone can exceed the pressure of the fluid by several orders of magnitude. 
In these regions, relatively small truncation errors in the evolution of the conserved variables may lead to large errors in the computation of the (internal) energy density and subsequently in the other primitive variables. 
This could in principle lead to the occurrence of regions with negative pressure or density. 
However, in such highly magnetized regions, \texttt{BHAC-QGP} is able to use the evolution equation for the entropy density $s$. 
In Milne coordinates, this evolution equation reads
  \begin{equation}
    \dfrac{\partial \mathcal{S}}{\partial \tau} + \partial_i \left(v^{i} \mathcal{S}\right) = -\dfrac{\mathcal{S}}{\tau}   \;, \label{EntropySwitch1}
  \end{equation}
where $\mathcal{S} := \gamma s$ is the entropy density in the Eulerian frame. 
Using the advection equation for the entropy one can discard the conservation of energy and sets the energy density to the value that is consistent with the entropy. 
Using the equation of state for a classical massless gas, the rescaled energy density will be then replaced by
  \begin{align}
    e^{\prime}_{\text{tot}} &= \xi - \left[\dfrac{\pi^{2/3} \left(\hbar c\right)}{g^{1/3}} \left(\dfrac{s}{4}\right)^{4/3}\right]  \notag \\[4pt]
    &\qquad 
    + \dfrac{1}{2} \left[\boldsymbol{B}^2 \left(1 + \boldsymbol{v}^2\right) - \left(\dfrac{\boldsymbol{S} \cdot \boldsymbol{B}}{\xi}\right)^2\right]  \;, \label{EntropySwitch2}
  \end{align}
where the second term on the right-hand side represents the pressure calculated from the entropy density. 
In this way, problematic regions where negative pressures occur can be eliminated. 
Using the advected entropy has also the advantage that the recovery procedure becomes independent of the energy density.
  \subsection{Adaptive Mesh Refinement}
The computational grid employed in \texttt{BHAC} and thus also in \texttt{BHAC-QGP} is provided by the \texttt{MPI-AMRVAC} toolkit. 
Adaptive Mesh Refinement (AMR) is a powerful numerical technique which allows one to refine only the regions of interest. 
Instead of using a high resolution for the full computational domain, only those areas undergoing significant changes are calculated with a high spatial resolution. 
This is computationally very efficient, since far fewer cells with high resolution have to be computed. 
\par
AMR in \texttt{BHAC-QGP} comes with a fully adaptive block-based tree with a fixed refinement factor of two between successive levels for a simple implementation of parallelization. 
In essence, the computational domain in \texttt{BHAC-QGP} is first split into a number of blocks with an equal number of cells. 
The starting block is called root block, while blocks with higher resolution are called leaf blocks, so that a block-based ``tree'' is formed. 
If the root block is too coarse, it will spawn two leaf blocks in 1D, four leaf blocks in 2D, and eight leaf blocks in 3D. 
In this way, AMR allows for an effective solution of problems where it is necessary to resolve large gradients on small scales, but only in certain regions of the computational domain. 
Which zones of the mesh need to be refined are determined with the help of a mesh-refinement criterion. 
Basically, whenever a specified function of the conserved variables, e.g., the second or higher derivative, exceeds a prescribed threshold, the block is triggered for refinement. 
However, if this criterion parameter is set too small, then the computational costs are very high because too many areas are refined. 
On the other hand, if the parameter is too large, the relevant zones are not detected and not refined, making the algorithm inefficient. 
In \texttt{BHAC-QGP}, the refinement is triggered in a completely automated way using the Löhner scheme \cite{Lohner:1987cfd}. 
For more details we refer to Chapter 2.11 of the original \texttt{BHAC} paper \cite{Porth:2016rfi}.
\section{Numerical Tests}
\label{NumericalTests}
In order to prove the viability of the implementation of the RMHD equations in Milne coordinates and the ultrarelativistic EOS we present in the following some tests. 
If not stated otherwise, we use ``minmod'' as reconstruction scheme and a third-order Runge-Kutta scheme (RK3) for time stepping. 
  \subsection{Bjorken flow}
The Bjorken flow describes the longitudinal expansion of a fluid whose velocity in Milne coordinates is $\tilde{u}^\mu = \left(1, 0, 0, 0\right)^T$ \cite{Bjorken:1982qr}. 
With the ultrarelativistic EOS and the evolution equation for the magnetic field (\ref{MilneCoords19}), the conservation equation for the energy (\ref{MilneCoords12}) simplifies to \cite{Roy:2015kma}
  \begin{equation}
    \dfrac{\partial \overline{e}}{\partial \tau} + \dfrac{4}{3} \dfrac{\overline{e}}{\tau} = 0 \;,  \label{BjorkenFlow1}
  \end{equation}
where $\overline{e} := e\!\left(\tau\right)/e\!\left(\tau_0\right)$. 
The evolution of the energy density is therefore given by
  \begin{equation}
    \overline{e}\left(\tau\right) = \left(\dfrac{\tau_0}{\tau}\right)^{4/3} \;,   \label{BjorkenFlow2} 
  \end{equation}
which states that the energy density decays with time as a result of the expansion of the fluid. 
Although a magnetic field was included, the energy density still follows the usual Bjorken expansion law because the ratio $\boldsymbol{B}/e^{3/4}$ is conserved in ideal MHD due to the frozen-flux theorem \cite{Roy:2015kma}.
\par
\begin{figure}[!ht]
    \centering
    \includegraphics[width=0.94\linewidth]{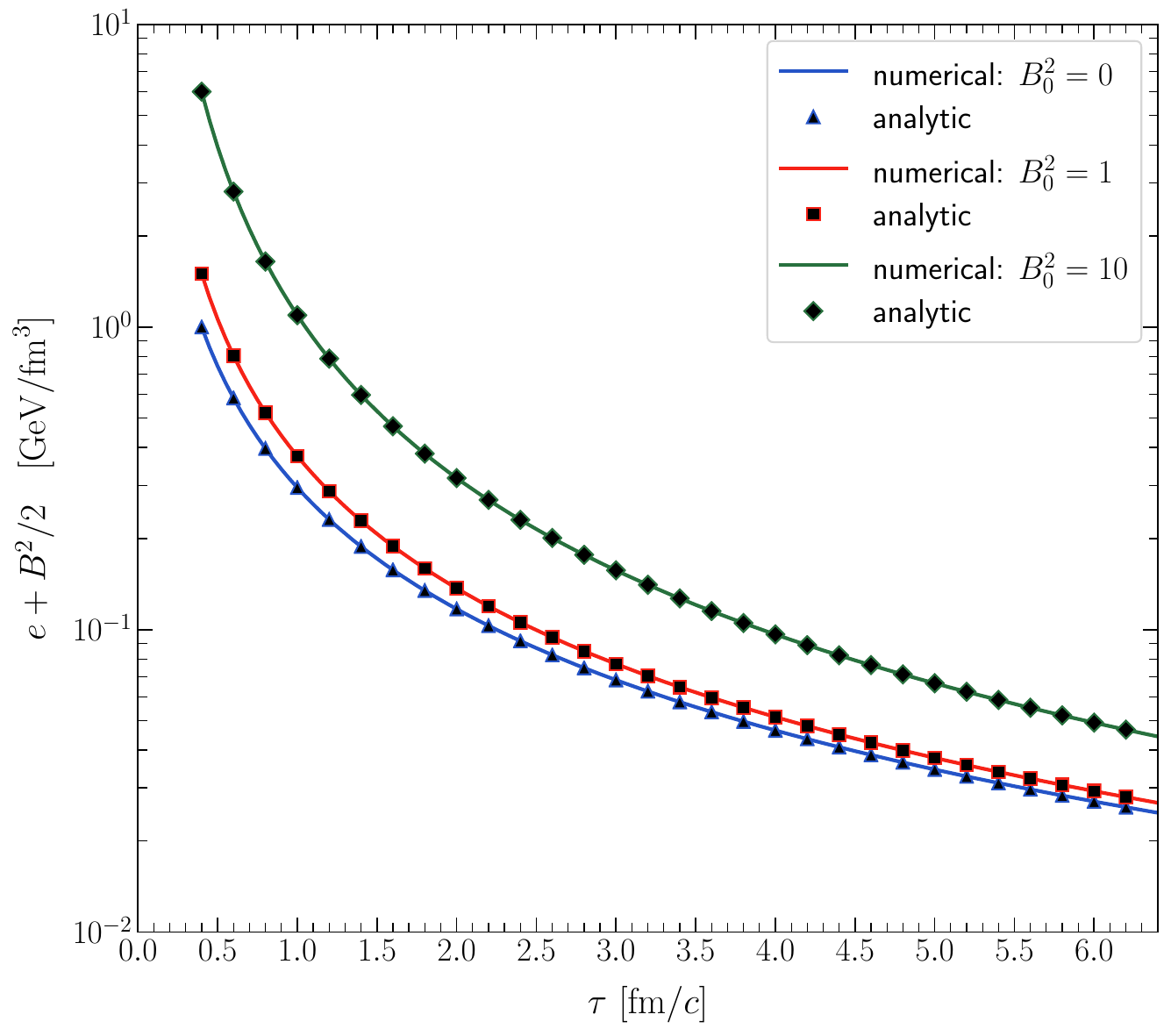}
    \caption{Temporal evolution of the total energy density for different values of the initial magnetic field. The results obtained from \texttt{BHAC-QGP} (solid lines) show a perfect agreement with the analytic solution (symbols).}
    \label{Fig:BjorkenFlow1}
  \end{figure}

We calculated the Bjorken flow with \texttt{BHAC-QGP} for three different values of the initial magnetization, $\boldsymbol{B}_0^2 = \left\{0; 1 \,\mathrm{GeV/fm^3}; 10\,\mathrm{GeV/fm^3}\right\}$. 
From Fig.\ \ref{Fig:BjorkenFlow1} one observes that the numerical results show a perfect agreement with the analytic solution at any time $\tau$. 
As mentioned above, the magnetic field lines are simply advected with the fluid and a stronger magnetic field only contributes correspondingly more to the total energy density. 
Nevertheless, the strength of the field has no influence on the temporal evolution of the energy density.
  \subsection{Gubser flow}
An extension of the Bjorken flow is the so-called Gubser flow \cite{Gubser:2010ze}, which is a special solution for an azimuthally symmetric and longitudinally boost-invariant flow profile, i.e., it only depends on $\tau$ and the radial coordinate in the transverse plane, $r:= \sqrt{x^2 + y^2}$. 
Gubser's solution reads
  \begin{align}
    e = \hat{e}_0 \left(\dfrac{4q^2}{\tau}\right)^{4/3}    \left[1 + 2  q^2  \left(\tau^2 + r^2\right) + q^4  \left(\tau^2 - r^2\right)^2\right]^{-4/3}\!\!, \label{GubserFlow1}
  \end{align}
where $\hat{e}_0$ is a normalization parameter for the energy density and $q$ is a parameter with the dimension of an inverse length. 
In addition, the components of the four-velocity are:
  \begin{eqnarray}   
    u^{\tau} &=& \cosh \kappa              \label{GubserFlow2} \;, \\[4pt]
    u^{x} &=& \dfrac{x}{r} \, \sinh\kappa  \label{GubserFlow3} \;, \\[4pt]
    u^{y} &=& \dfrac{y}{r} \, \sinh \kappa  \label{GubserFlow4} \;, \\[8pt]
    u^{\eta_S} &=& 0 \;,  \label{GubserFlow5}
  \end{eqnarray}
where 
  \begin{equation}
    \kappa\left(\tau, r\right) = \text{Artanh}\!\left(\dfrac{2  q^2  \tau  r}{1 + q^2  \tau^2 + q^2  r^2}\right) \;. \label{GubserFlow6}
  \end{equation}
Reproducing the Gubser flow profile provides a highly non-trivial theoretical benchmark for any numerical code solving hydrodynamics.
\par
\begin{figure*}[!ht]
    \centering
    \includegraphics[width=0.47\linewidth]{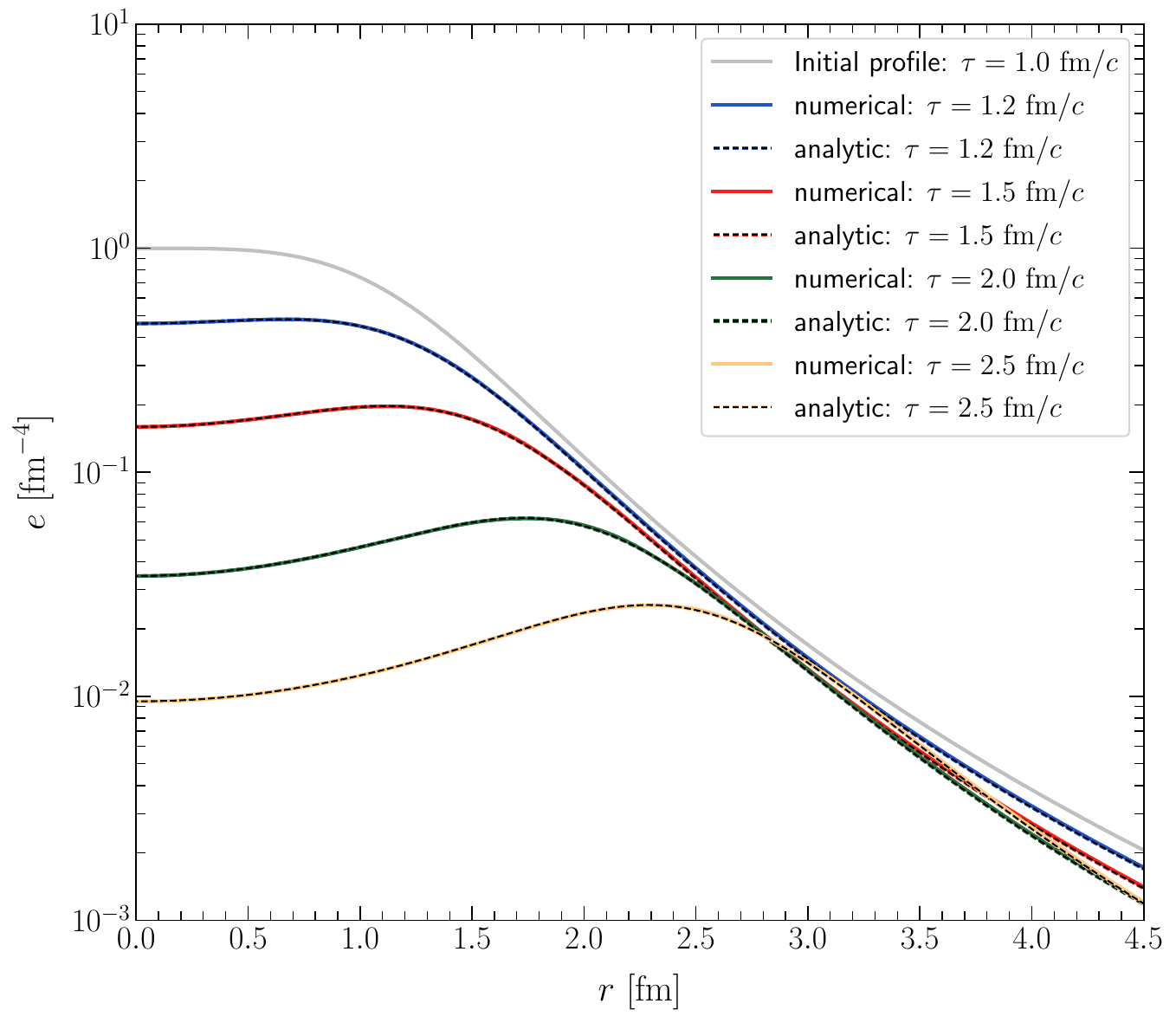}\hfill
    \includegraphics[width=0.47\linewidth]{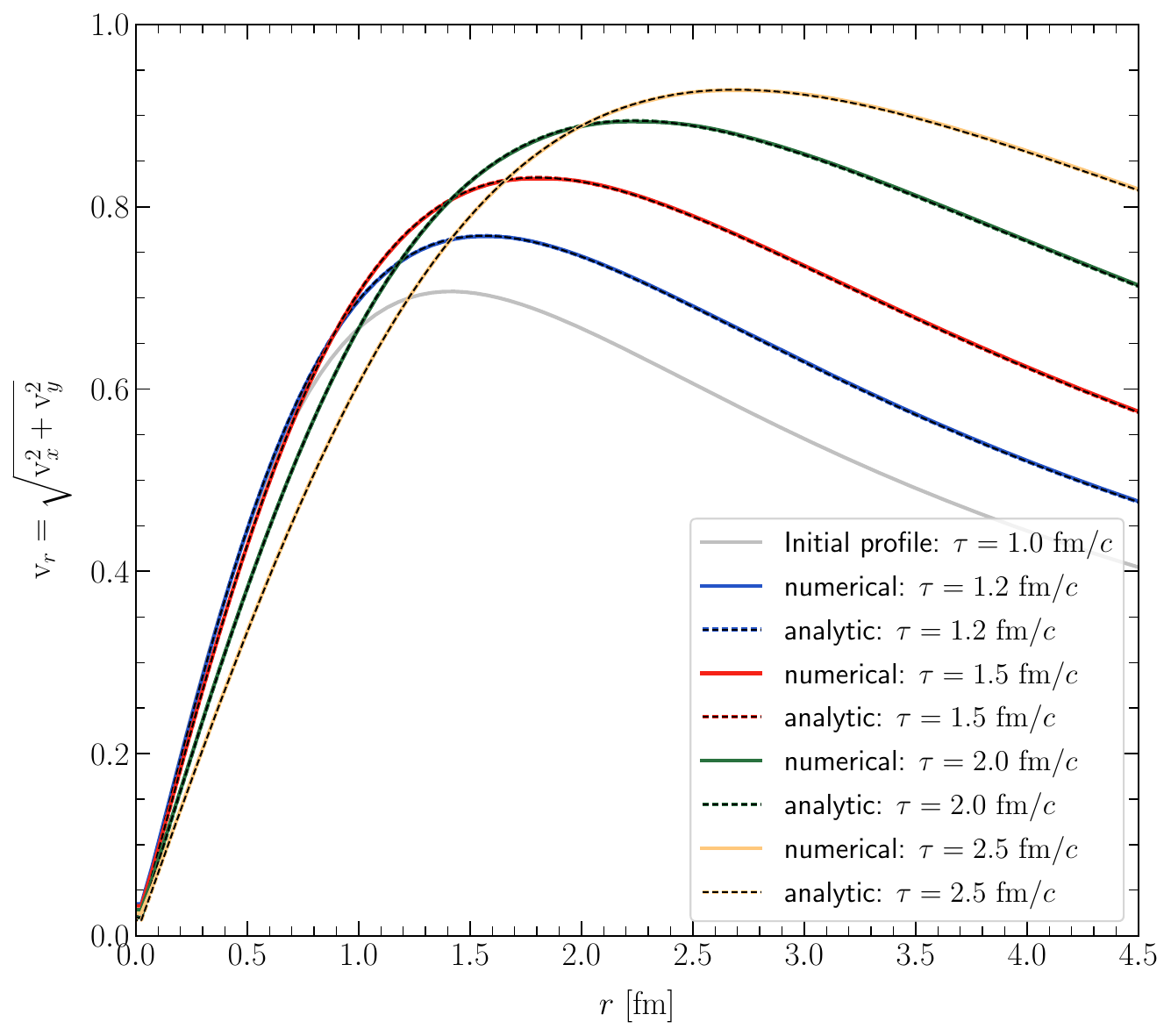}
    \caption{Radial dependence of the energy density (left panel) and of the radial velocity (right panel) at different times. The results obtained from a (2+1)-dimensional simulation in \texttt{BHAC-QGP} (solid lines) show a perfect agreement with the analytic solution (dashed lines).}
    \label{Fig:GubserFlowFig1}
  \end{figure*}

In order to perform this test in \texttt{BHAC-QGP}, we initialize the energy density according to Eq.\ (\ref{GubserFlow1}) at time $\tau_0 = 1~\text{fm}$ with $q = 1~\text{fm}^{-1}$ and $\hat{e}_0 = 1$. 
In the same way, we initialize the velocity profile according to Eqs.\ (\ref{GubserFlow2})--(\ref{GubserFlow5}). 
The size of the computational domain is $\left[-10.0~\mathrm{fm}, 10.0~\mathrm{fm}\right] \times \left[-10.0~\mathrm{fm}, 10.0~\mathrm{fm}\right]$ with $400 \times 400$ cells.
The comparisons between the numerical (solid lines) and the analytic solutions at different time steps are presented in Fig.\ \ref{Fig:GubserFlowFig1}. 
\texttt{BHAC-QGP} reproduces the analytic solution (dashed lines) for the energy density and for the radial velocity $\mathrm{v}_r = \left(v_x^2 + v_y^2\right){}^{1/2}$ perfectly.

\subsection{Hydrodynamic freeze-out with semi-analytical Gubser flow}
 In the previous section we have demonstrated that \texttt{BHAC-QGP} reproduces the analytic Gubser solution perfectly. 
 We now compute the isothermal hypersurfaces for the corresponding flow profile and calculate the particle spectra on the freeze-out hypersurface, where the temperature assumes a certain value $T_f$. 

The single-inclusive momentum spectrum for hadrons of species $i$ is then given by \cite{Cooper:1974mv}
\begin{equation}\label{CooperFrye1}
 p_i^0 \frac{\mathrm{d} N_i}{\mathrm{d}^3p} = \frac{g_i}{(2\pi)^3} \int \mathrm{d}\Sigma_\mu \, p_i^\mu f\left(-\frac{p_i^\mu u_\mu}{T_f}\right)\;,
\end{equation}
where 
$p_i^\mu = (p_i^0, \boldsymbol{p})^T$ is the on-shell four-momentum of a hadron of species $i$, where the energy is $p_i^0 = \sqrt{\boldsymbol{p}^2 + m_i^2}$, $\mathrm{d}\Sigma_\mu$ is the directed area element normal to the freeze-out hypersurface, $g_i$ is the degeneracy factor of hadron species $i$, and $f(x)$ is a distribution function that we will take as the Boltzmann distribution $f(x) = \exp(-x)$.  
(The sign in the argument of $f$  comes from our use of the mostly + signature metric.). 
We choose Boltzmann rather than Fermi-Dirac or Bose-Einstein statistics to be consistent with our assumptions regarding the equation of state, see Sec.\ \ref{subsec:EoS}.
An additional advantage is that most subsequent calculation can be done analytically.

The freeze-out surface for Gubser flow can be parametrized as
\[
\Sigma^\mu = (\tau_f(r), \eta_S, r, \phi)^T\;,
\]
where $\tau_f(r)$ is the solution to the equation $T(r, \tau_f) = T_f$. 
This equation can be solved using standard root-finding algorithms. 
In our computations, we used \texttt{Mathematica}'s \textit{NSolve} function, which provides a robust tool for solving non-linear equations numerically \citep{Mathematica}.

  \begin{figure}[ht]
    \centering
    \includegraphics[width=1.0\linewidth]{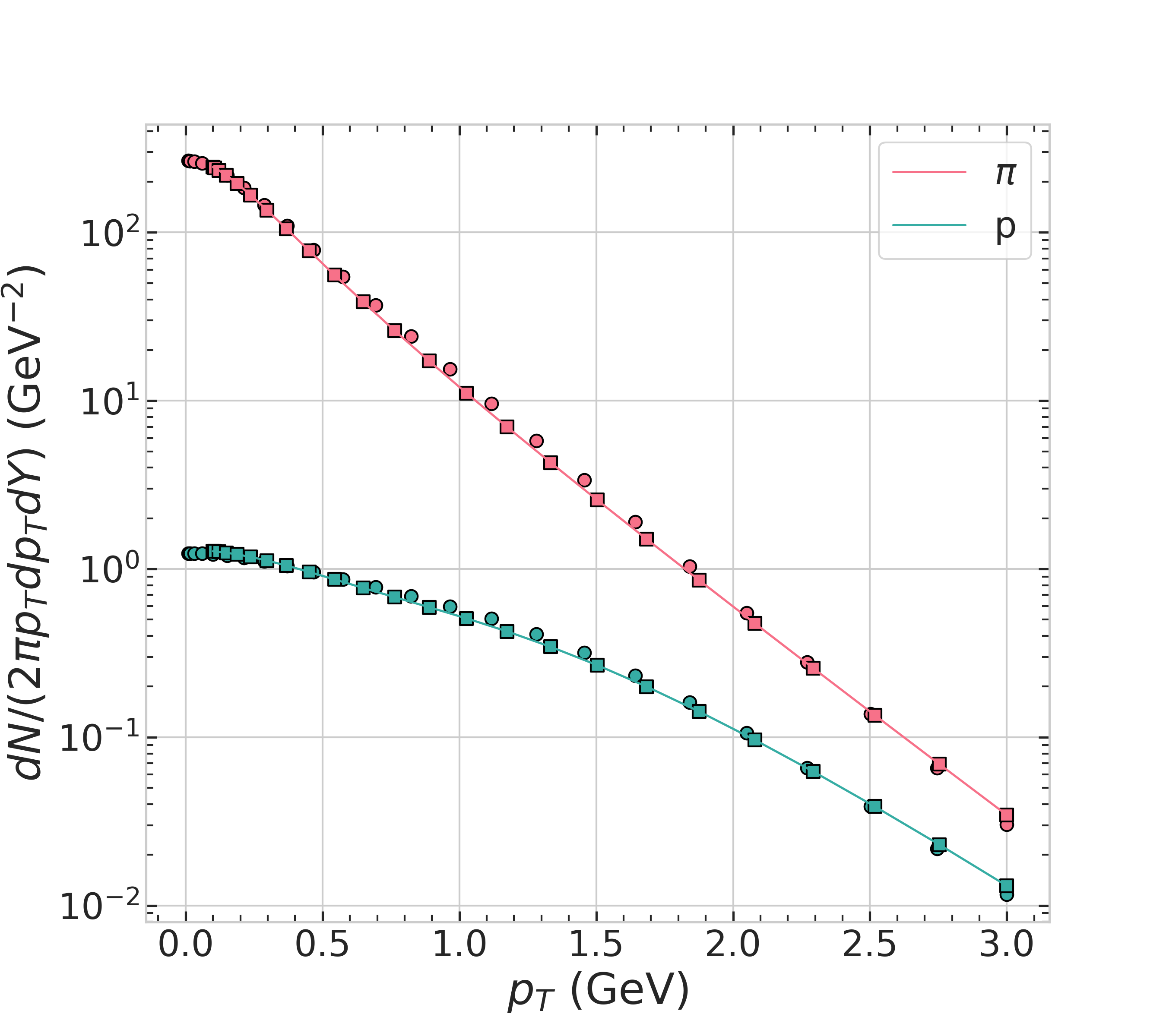}
\caption{Comparison of the spectrum of positively charged pions and protons as a function of transverse momentum $p_T$ resulting from Gubser’s hydrodynamic solution. 
Solid lines represent the semi-analytic solution obtained from Eq.\ \eqref{Eq:AnalyticalGubser}, while symbols (squares for 2+1 dimensions and circles for 3+1 dimensions) represent the results of numerical simulations from \texttt{BHAC-QGP}.}
    \label{Fig:AnalyticalGubser}
\end{figure}

The directed area element normal to the freeze-out surface is \begin{align}
    \mathrm{d}\Sigma_\mu & = \epsilon_{\mu\nu\lambda\rho} \frac{\partial \Sigma^\nu}{\partial \eta_S} \frac{\partial \Sigma^\lambda}{\partial r} \frac{\partial \Sigma^\rho}{\partial \phi} \sqrt{-g} \, \mathrm{d}\eta_S \, \mathrm{d}r \, \mathrm{d}\phi\nonumber \\
   & = (-1, 0, -R_f, 0) \, r \,\tau_f (r)\, \mathrm{d}\eta_S \, \mathrm{d}r \, \mathrm{d}\phi\;,
\end{align}
where $\sqrt{-g} = r \, \tau_f(r)$ on the freeze-out surface. 
We have used the fact that $\mathrm{d}T = \frac{\partial T}{\partial r}\, \mathrm{d}r + \frac{\partial T}{\partial \tau}\, \mathrm{d}\tau = 0$ on the freeze-out surface to define
\begin{equation}
    R_f := -\frac{\partial \tau_f(r)}{\partial r} = \frac{\partial T_f/\partial r}{\partial T_f/\partial \tau_f}\;.
\end{equation}

We choose $\hat{e}_0 = 13605$, $q^{-1} = 6.4$~fm, and $T_f = 0.13$~GeV, respectively. 
(The value of $q$ is changed as compared to that used in the previous section in order to obtain a more realistic shape of the momentum spectrum.)
With these specifications, we can now compute the spectrum \eqref{CooperFrye1} for hadrons of species $i$.

For $ u^\mu $ as in Gubser's flow, the argument of the function $ f $ simplifies to
\begin{equation}
    p_i^\mu u_\mu = -m_{iT} u^\tau \cosh(Y - \eta_S) + p_T u^r \cos(\phi_p - \phi)\;,
\end{equation}
where $ m_{iT} := \sqrt{p_T^2 + m_i^2}$ is the transverse mass of hadron species $i$, $ p_T $ is the transverse momentum, $Y := \frac{1}{2} \ln [(p_i^0 + p^z)/(p_i^0 - p^z)]$ is the longitudinal rapidity of the particle, $ \phi_p $ is the azimuthal angle of the momentum of the particle, and $\phi $ is the azimuthal angle of the three-velocity of the fluid.

The $\eta_S$ and $\phi$ integrals in Eq.\ \eqref{CooperFrye1} can be performed analytically, yielding
\begin{align}
    p_i^0 \frac{\mathrm{d} N_i}{\mathrm{d^3}p} & = \frac{g_i}{2\pi^2} \int \mathrm{d}r \, r \, \tau_f (r) \notag\\
    &\times \bigg[ m_{iT}K_1 \left( \frac{m_{iT} u^\tau}{T_f} \right) I_0 \left( \frac{p_T u^r}{T_f} \right) \notag \\
    &\quad + R_f p_T K_0 \left( \frac{m_{iT} u^\tau}{T_f} \right) I_1 \left( \frac{p_T u^r}{T_f} \right) \bigg]\;, \label{Eq:AnalyticalGubser}
\end{align}
where $K_n$ and $I_n$ are modified Bessel functions. 
Since the solution is boost-invariant in longitudinal direction and cylindrically symmetric in transverse direction, the result is independent of $Y$ and $\phi_p$ and depends only on $p_T$. 
The radial $r$ integral on the freeze-out surface is evaluated numerically to obtain the single-inclusive momentum spectra of hadrons freezing out from Gubser’s hydrodynamic flow.

To validate the correctness of the \texttt{BHAC-QGP} simulation coupled with the \texttt{CORNELIUS} method for finding the hypersurface \cite{Huovinen:2012is}, we initialized the energy density and fluid velocities corresponding to the parameters discussed above at time $\tau_0 = 0.4$~fm. 
We performed both (2+1)-dimensional and (3+1)-dimensional simulations: (i) For the (2+1)-dimensional simulation, we defined the computational domain in the range $[-30\, \mathrm{fm}, 30\, \mathrm{fm}]^2$ 
with 200 cells in each direction, and (ii) for the (3+1)-dimensional simulation, we defined the computational domain in the range $[-30\, \mathrm{fm}, 30\, \mathrm{fm}]^2 \times [-30,30]$ with $300 \times 300 \times 100$ cells in the $(x, y, \eta_S)$ directions. 
In the rapidity direction for the (3+1)-dimensional simulation, we assumed a flat profile to mimic boost invariance. 
The magnetic field was turned off in these simulations.

Figure \ref{Fig:AnalyticalGubser} shows the spectrum of positively charged pions and protons as a function of transverse momentum $p_T$ resulting from Gubser’s hydrodynamic solution. 
Solid lines represent the semi-analytic solution obtained from Eq.\ \eqref{Eq:AnalyticalGubser}, while symbols (squares for 2+1 dimensions and circles for 3+1 dimensions) represent the results of numerical simulations from \texttt{BHAC-QGP}. 
We observe perfect agreement between the semi-analytical solution and \texttt{BHAC-QGP} for the (2+1)-dimensional simulations, and near-perfect agreement for the (3+1)-dimensional simulations.

  \subsection{Relativistic fluid expansion into vacuum}
 In this section, we discuss (i) the one-dimensional expansion into vacuum and (ii) the cylindrically symmetric transverse expansion with boost-invariant longitudinal expansion into vacuum.
 In case (i), we assume the fluid to have a finite extension $2R$ in the (single) spatial direction, while in case (ii) we assume the diameter of the transverse extension of the fluid to be $2R$. 
 Since in case (i) the system is symmetric around the origin, it is sufficient to consider the solution in the positive half-plane.
 In case (ii), because of the longitudinal boost invariance we can restrict the consideration to the ($z=0$)-plane, where $\tau \equiv t$.
 Thus, the initial condition at time $t_0$ is in both cases
\begin{equation}\label{Eq:FiniteInitEn}
e(r, t_0) =
\begin{cases}
e_0, & \text{if } r \le R \\
0, & \text{if } r > R
\end{cases}\;,
\end{equation}

\begin{equation}\label{Eq:FiniteInitvr}
v(r, t_0) =
\begin{cases}
0, & \text{if } r \le R \\
1, & \text{if } r > R 
\end{cases}\;,
\end{equation}
where $r\geq 0$ is the spatial variable in case (i) and the transverse radial variable in case (ii).
This setup has been thoroughly investigated by various authors, particularly Baym et al.\ \cite{Baym:1983amj}. 

\subsubsection{One-dimensional expansion}

In case (i), a completely analytical expression for the space-time evolution of the fluid is possible by matching the Riemann rarefaction waves, which travel with the velocity of light into the vacuum at the left and right edges of the system, to the Landau-Khalatnikov solution \cite{Landau:1953wku,Khalatnikov:1954}, which is valid in the region where the two rarefaction waves travelling inwards from the two edges of the system overlap. 
Nevertheless, we choose a much simpler method to construct the solution, which also applies in case (ii), namely by using the method of characteristics to recast the coupled partial differential equations for the hydrodynamic variables into a set of coupled ordinary differential equation \cite{Rischke:1995ir}.

\begin{figure*}[!ht]
    \centering
    \includegraphics[width=0.46\linewidth]{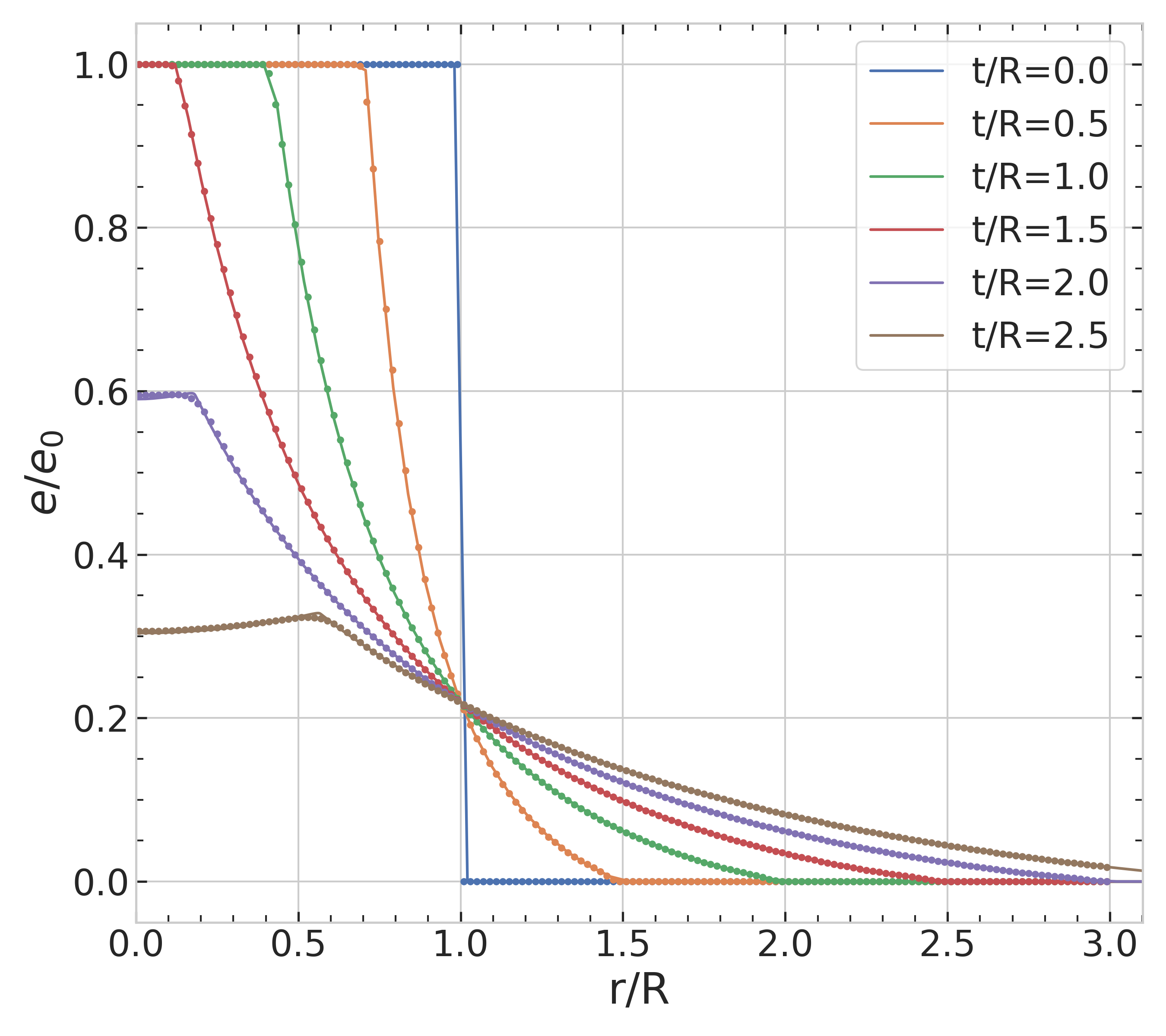}\hfill
    \includegraphics[width=0.46\linewidth]{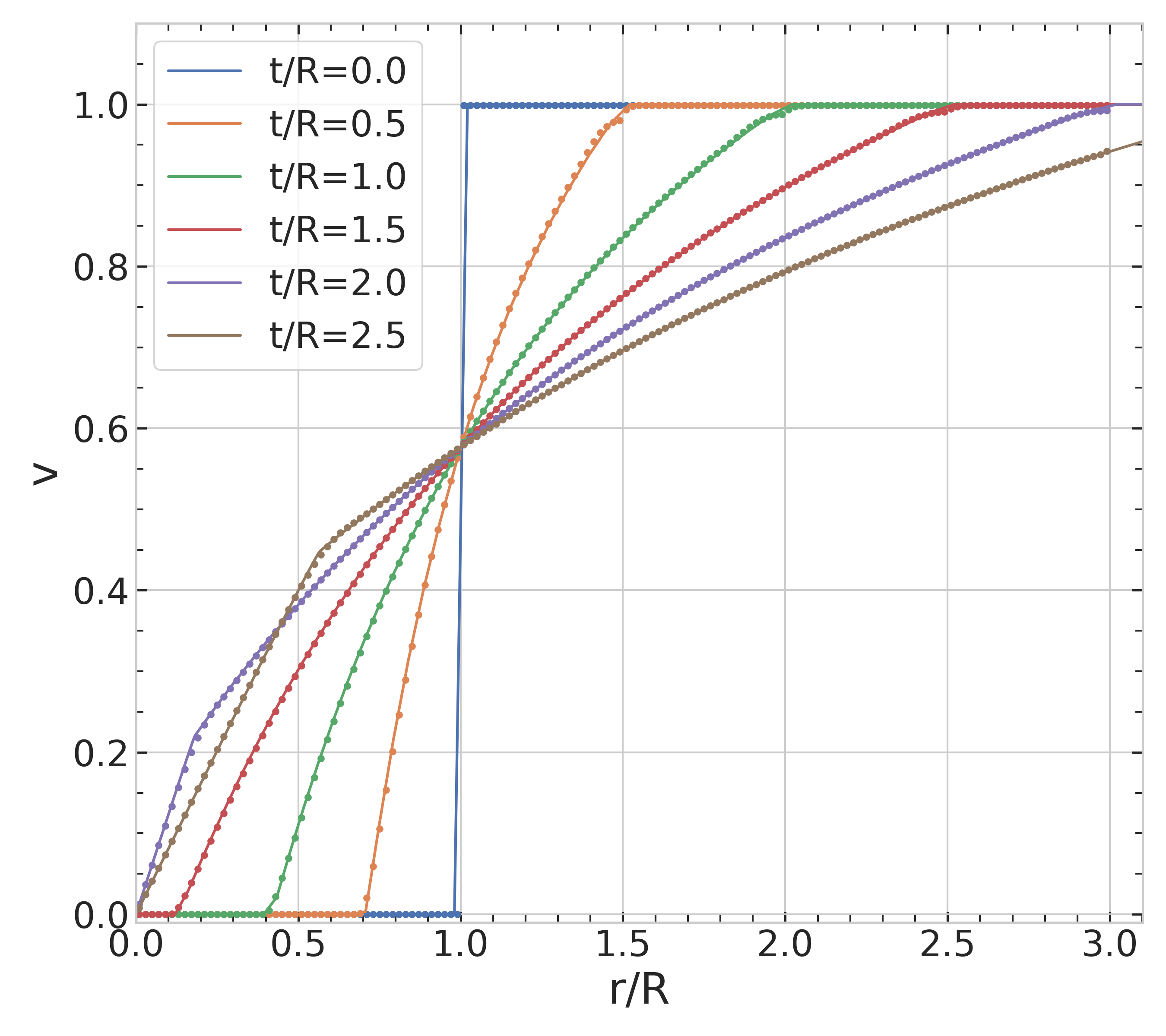}
    \caption{Scaled energy-density (left panel) and velocity (right panel) profiles at different times for one-dimensional expansion into vacuum. 
    The (2+1)-dimensional results from \texttt{BHAC-QGP} (symbols) are in good agreement with those obtained from method of characteristics (solid lines).}
    \label{Fig:Finitematter}
  \end{figure*}

Figure~\ref{Fig:Finitematter} (left and right panels) shows the spatial distribution of the energy density
normalized to its value $e_0$ at $t_0$ and of the velocity at various times for case (i). 
The lines are obtained from the method of characteristics while the symbols denote the results from \texttt{BHAC-QGP}. 
For the numerical setup, we used a 2-dimensional Cartesian grid of size $[-3R,3R]\times [-3R,3R]$ with $200\times 200$ cells.
The $x$-direction is identified with the spatial variable $r$.
The initial conditions are assumed to be translationally invariant in $y$-direction.
Since for this case the hydrodynamic equations are invariant under the scale transformation $r\rightarrow \lambda r$ and $t\rightarrow \lambda t$, the distance and time scales are determined by $R$, which we set to $R=1$.
\par
A few salient points about the time evolution are noted:
\begin{enumerate}
\item Without loss of generality, we may set $t_0=0$.
\item A Riemann rarefaction wave \cite{Baym:1983amj} travels to the left from the right edge of the system with the velocity of sound $c_s = 1/\sqrt{3}$.
Thus, for times $t/R \leq 1/c_s = \sqrt{3}$, the matter at space points
$|r|/R \leq 1/\sqrt{3}$ remains undisturbed in the initial state (\ref{Eq:FiniteInitEn}), (\ref{Eq:FiniteInitvr}).
\item At $t/R = 1/c_s$, the rarefaction wave from the right edge of the system starts to overlap with that from the left edge. 
For later times, the Landau-Khalatnikov solution \cite{Landau:1953wku,Khalatnikov:1954} applies for small $r$, while for large $r$ the Riemann form persists. 
The two solutions are matched at a certain point, e.g., for $t/R=2.5$ near $r/R \sim 0.6$, where a maximum of the  energy density appears, cf.\ Fig.\ \ref{Fig:Finitematter} (left panel). 
The matching point between the two solutions appears as a kink in the velocity distribution in Fig.\ \ref{Fig:Finitematter} (right panel).
\item There exists a stationary point in both energy density and velocity distribution for $t/R < 1/c_s$, corresponding to $v_r = c_s$ and $e/e_0 \sim 0.21$. 
For late times (not shown here), when the fluid interior has cooled below $e/e_0 \sim 0.21$, the point stops being stationary.
\end{enumerate}
Overall, we see good agreement between the results from \texttt{BHAC-QGP} and those obtained using the method of characteristics.

\subsubsection{Cylindrically symmetric transverse expansion with longitudinal boost invariance}

Case (ii) is similar to the previous case. 
However, now there is a boost-invariant longitudinal expansion in the $z$-direction given by the well-known Bjorken solution $v_z = z/t$ \cite{Bjorken:1982qr} superimposed on the cylindrically symmetric transverse expansion. 
In this case we cannot set $t_0 =0$, but have to choose a value $t_0 >0$, as otherwise we would encounter singularities.
Therefore, for this setup, there are two characteristic length scales: the radius of the initial distribution $R$, and the time $t_0>0$ at which the initial conditions are specified.

\begin{figure*}[!ht]
    \centering
    \includegraphics[width=0.51\linewidth]{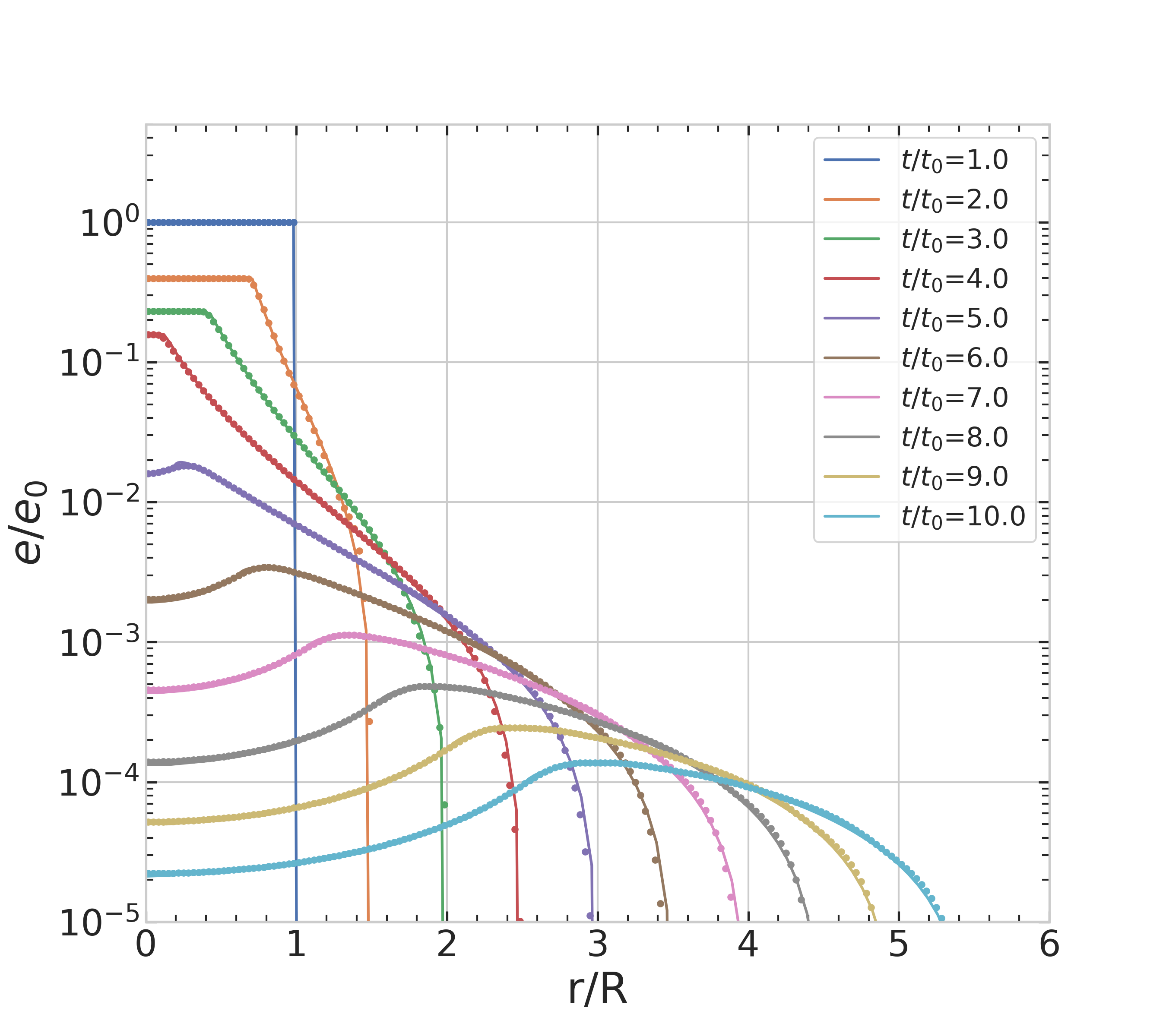}\hfill
    \includegraphics[width=0.46\linewidth]{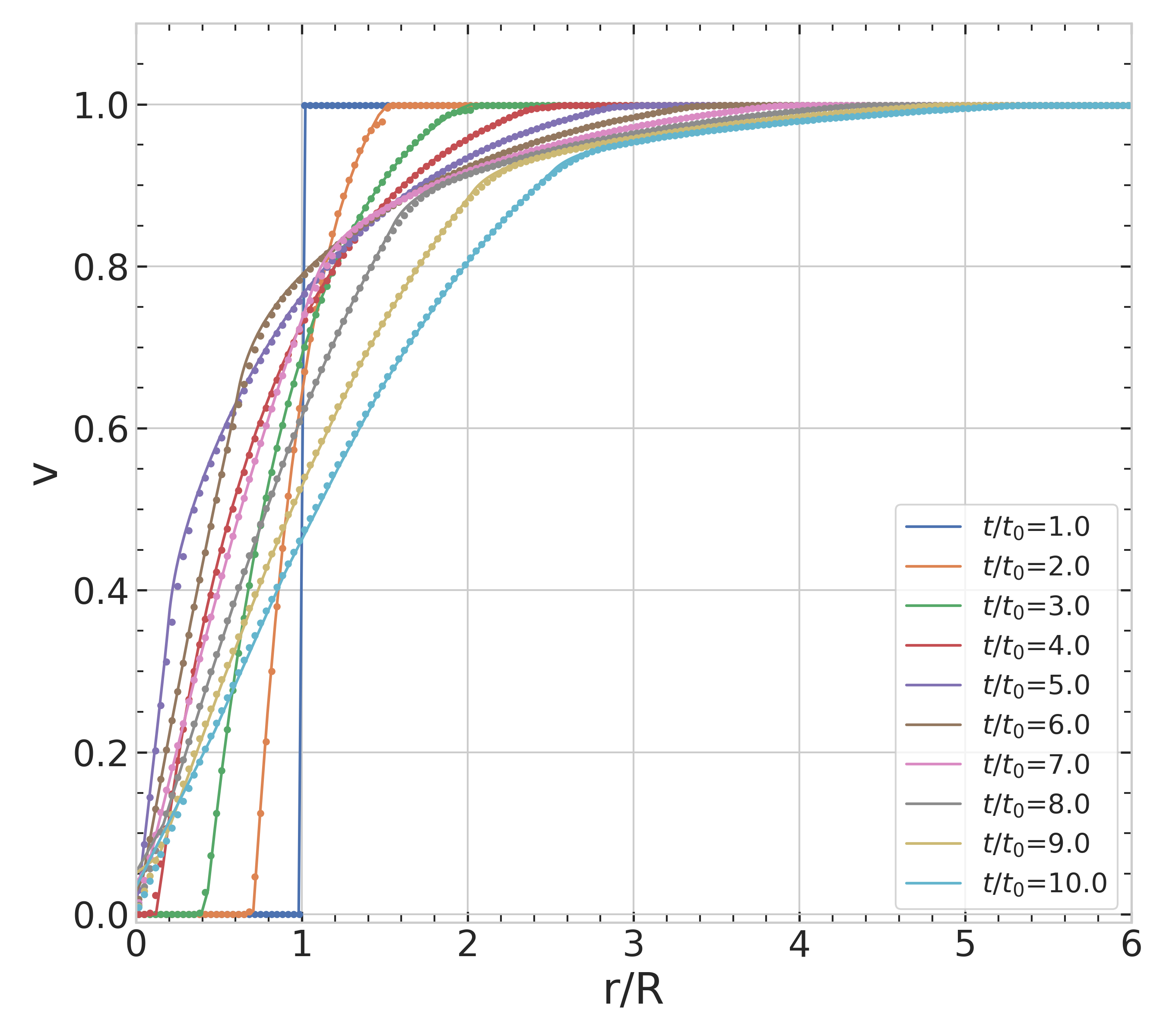}
    \caption{Scaled radial energy-density (left panel) and radial velocity (right panel) profiles at different times for the cylindrically transverse expansion with boost-invariant longitudinal expansion into vacuum. 
    The (2+1)-dimensional results from \texttt{BHAC-QGP} (symbols) are in good agreement with those obtained from method of characteristics (solid lines). }
    \label{Fig:Baym1}
  \end{figure*}

In Fig.\ \ref{Fig:Baym1} (left and right panels), we show the energy density normalized to its value $e_0$ at $t_0$ and the velocity as functions of the radial variable in the transverse plane at various times. 
The lines are obtained from the method of characteristics, while the symbols denote the results from \texttt{BHAC-QGP}. 
For the numerical setup, we used a 2-dimensional Cartesian grid of size $[-10R,10R]\times[-10R,10R]$ with $300\times300$ cells.
We set $R=1$ and $t_0=0.5$ (in arbitrary units). 
The radial energy-density and velocity distributions in Fig.\ \ref{Fig:Baym1} are plotted for $y=0$.

A few salient points about the time evolution are noted:
\begin{enumerate}
    \item Compared to the purely one-dimensional expansion, the present case also includes a coupling between the transverse and longitudinal motions.
    \item The longitudinal expansion causes a cooling of the fluid, as it spreads the entropy over a constantly increasing longitudinal interval. 
    The fluid cools uniformly at small $r$ even before the rarefaction waves overlap.
    \item Increasing $t_0$ decreases the cooling caused by the longitudinal expansion.
\end{enumerate}
Again, we see good agreement between the results from \texttt{BHAC-QGP} and those obtained using the method of characteristics.

  \subsection{Ultrarelativistic Orszag-Tang vortex}
\begin{figure*}[!htp]
    \centering
    \includegraphics[width=0.330\linewidth]{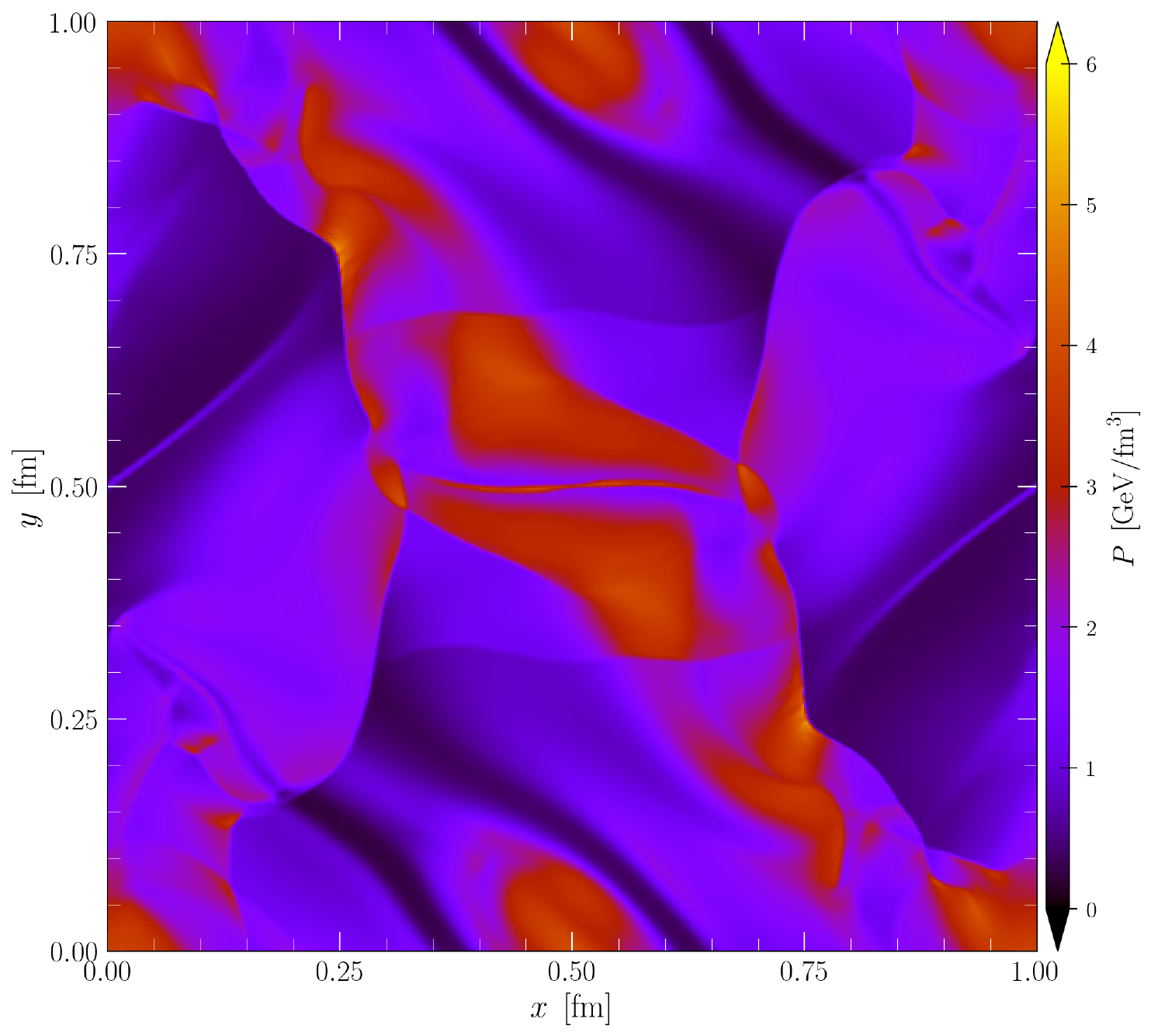}\hfill
    \includegraphics[width=0.330\linewidth]{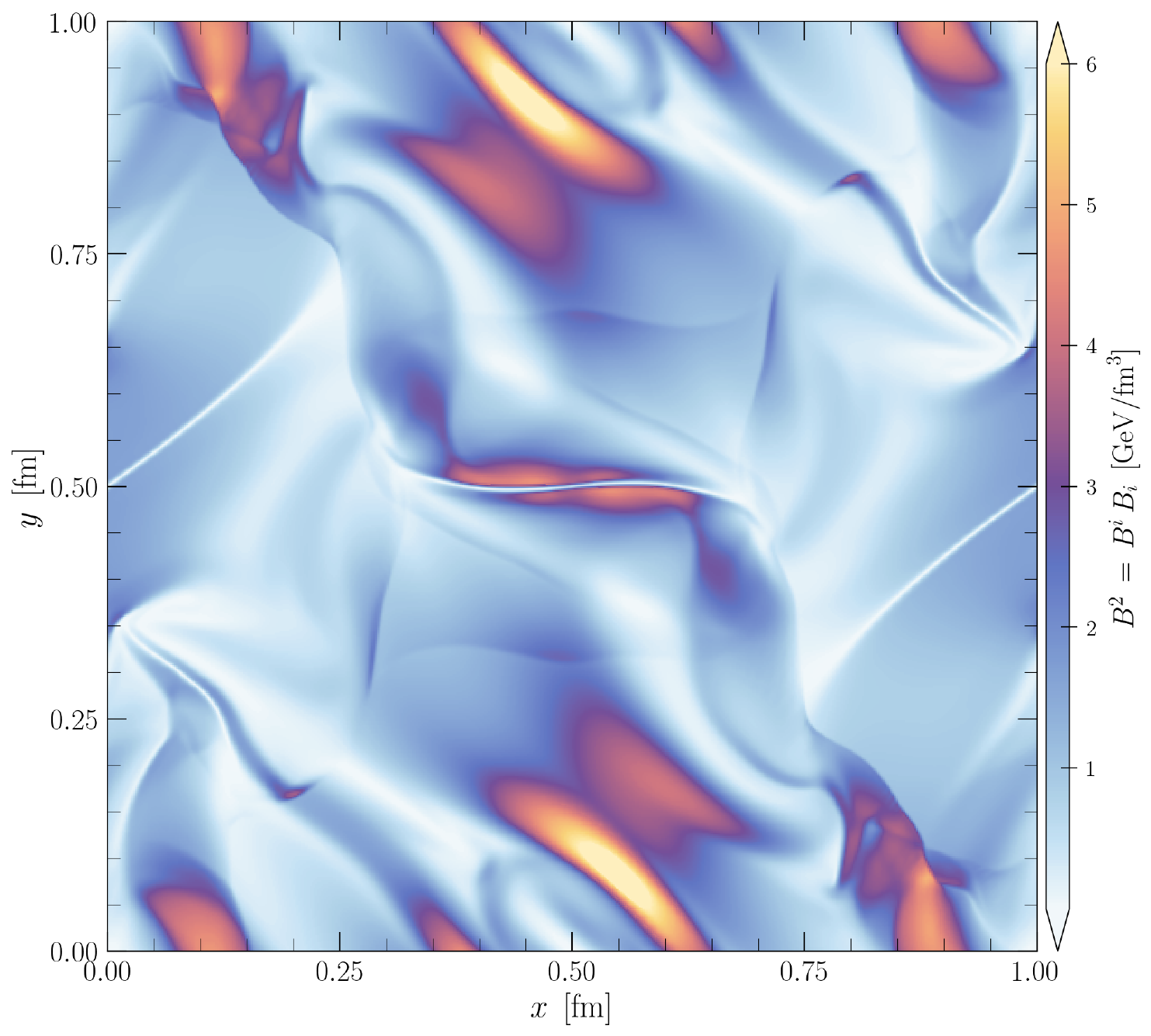}\hfill
    \includegraphics[width=0.330\linewidth]{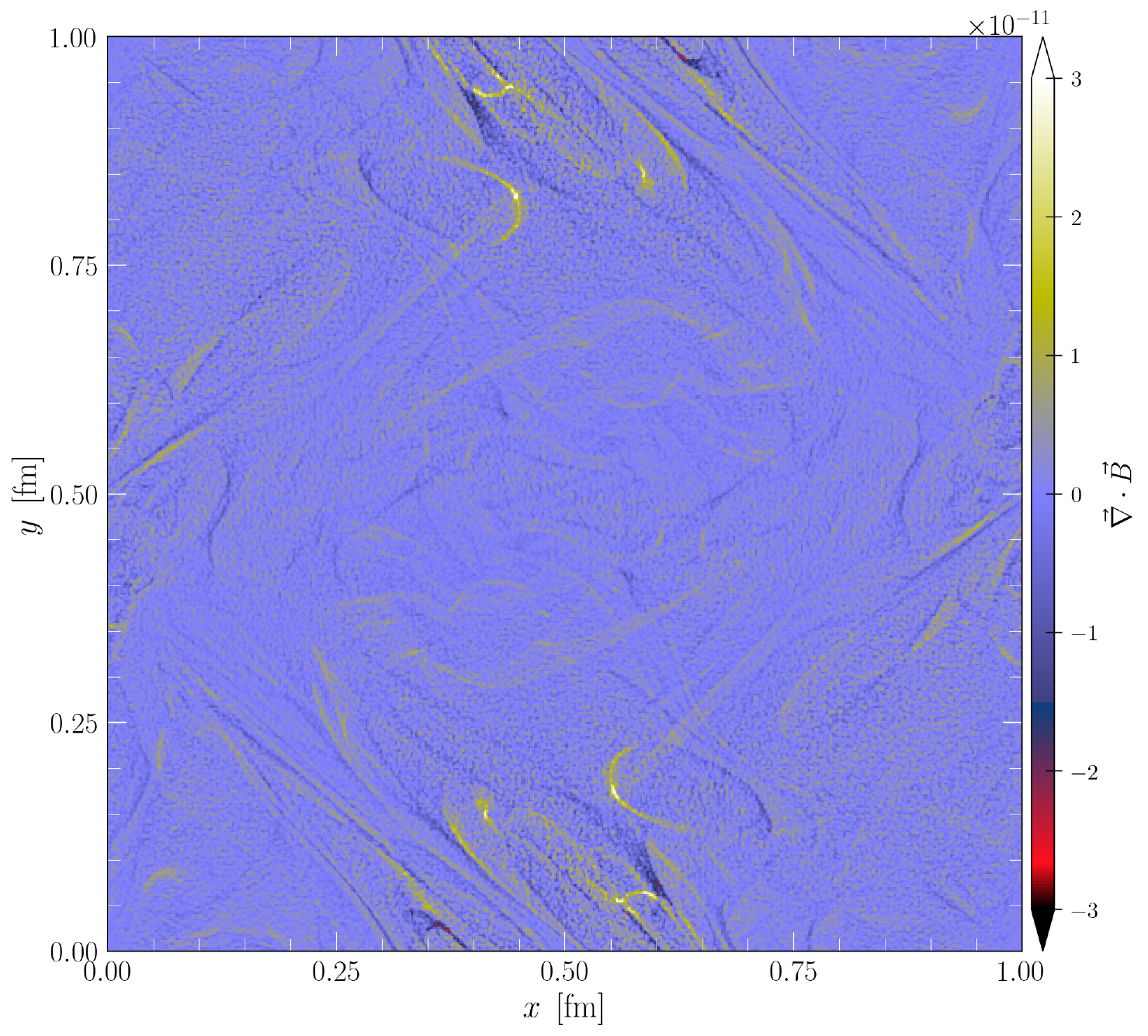} \\
    \includegraphics[width=0.330\linewidth]{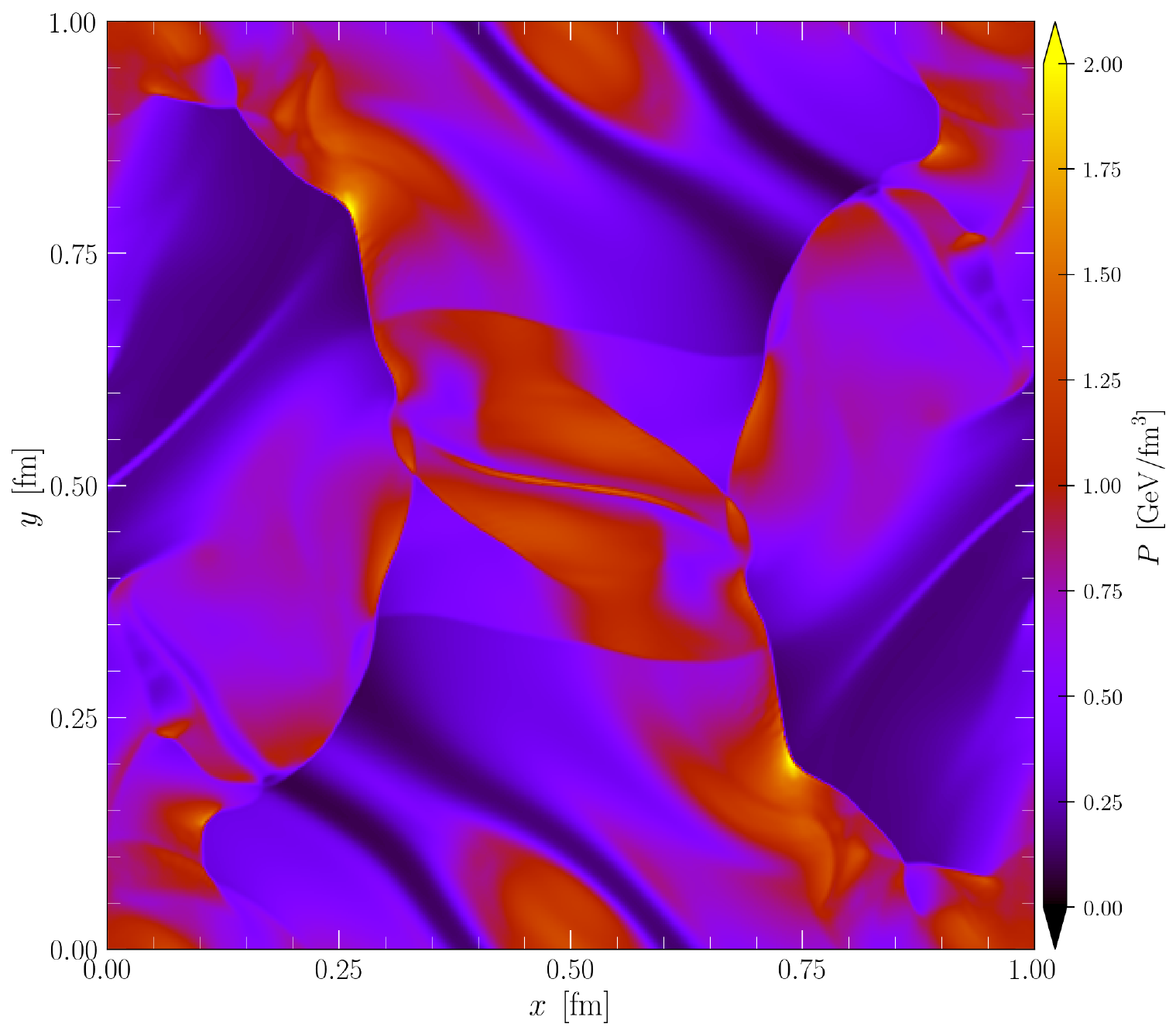}\hfill
    \includegraphics[width=0.330\linewidth]{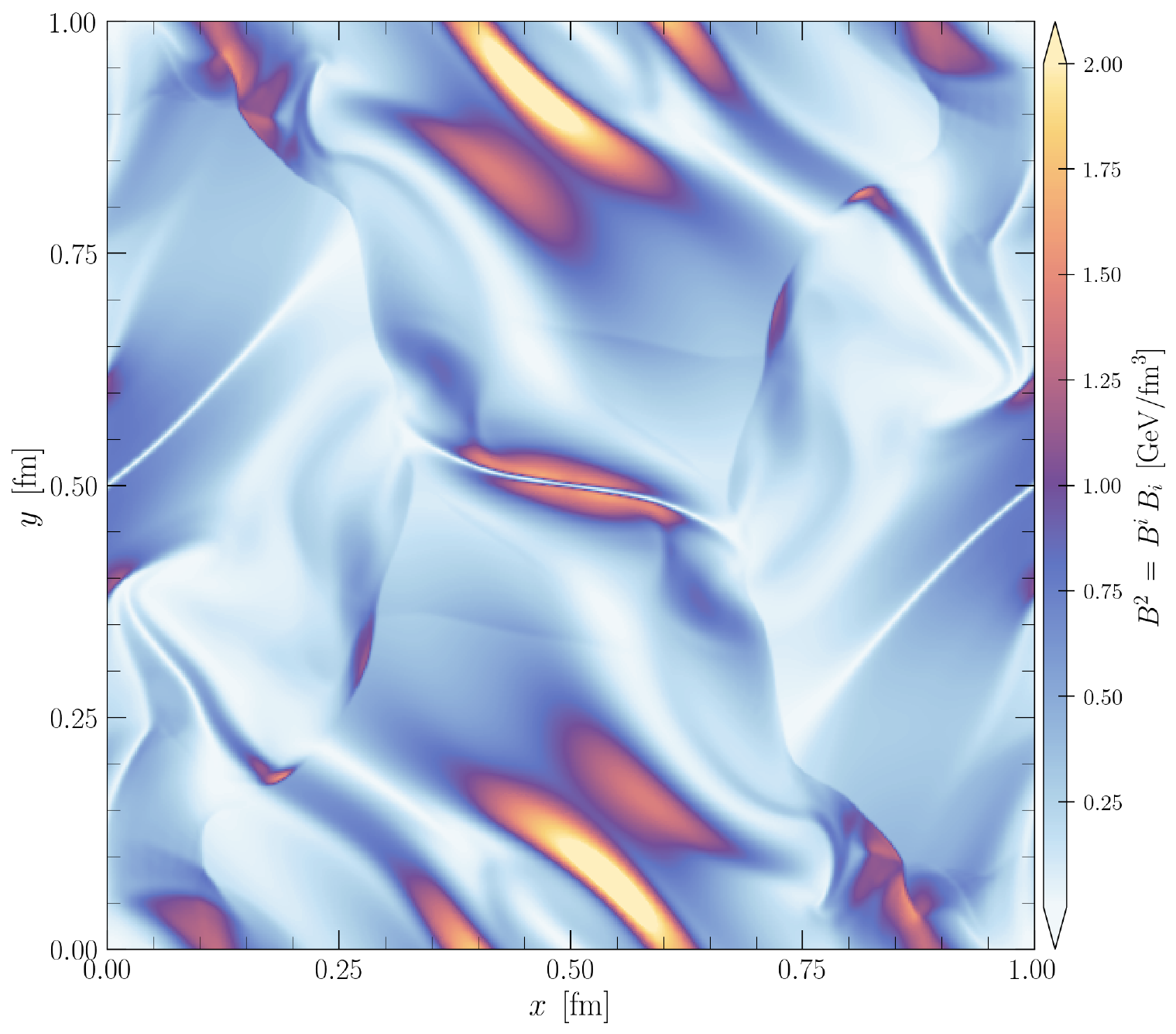}\hfill
    \includegraphics[width=0.330\linewidth]{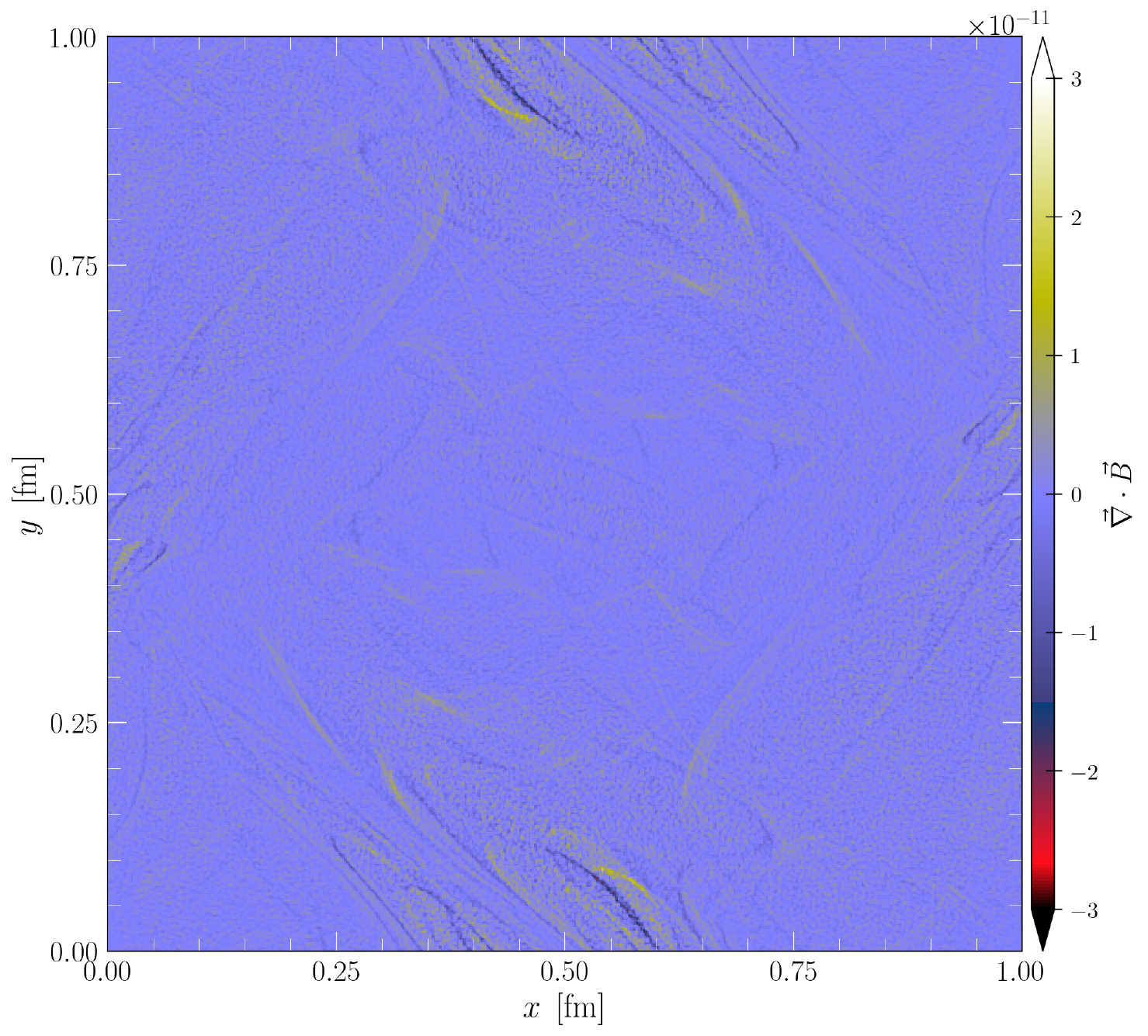}
    \caption{Quantities of the ultrarelativistic Orszag-Tang vortex test. 
    The first column shows the pressure $P$, the second column the magnetic field strength $B^i  B_i$, and the third column the violation of the divergence constraint $\boldsymbol{\nabla} \cdot \boldsymbol{B} = 0$. 
    The panels in the first row are obtained from a simulation in Cartesian coordinates at $t = 1~\mathrm{fm}$ (initial time $t_0 = 0~\mathrm{fm}$), while the second row shows the corresponding results of a calculation in Milne coordinates at $\tau = 2~\mathrm{fm}$ (initial time $\tau_0 = 1~\mathrm{fm}$). 
    The flow pattern is identical in both coordinates but the longitudinal expansion lead to a faster decay of the pressure and the magnetic field in Milne coordinates.}
    \label{Fig:OrszagTang1}
  \end{figure*}

A classical test for all MHD codes is the Orszag-Tang vortex problem \cite{Orszag:1979osz, Balsara:1999osz, Porth:2016rfi, Beckwith:2011iy, Stone:2008mh, Lopez-Miralles:2022ysg, Marcowith:2020vho}. 
This two-dimensional test is a useful problem to judge how robust the code is in handling the formation of shocks and its interactions. 
Furthermore, it can be used to test the divergence constraint, $\boldsymbol{\nabla} \cdot \boldsymbol{B} = 0$, and to check how strong magnetic monopoles affect the numerical solution. 
Although the Orszag-Tang test was already presented in the original \texttt{BHAC} code, we re-run it with the newly implemented EOS $P = e/3$ and in Milne coordinates with an initial time $\tau_0 = 1~\mathrm{fm}$. 
As computational domain we define a square of $\left[0.0~\mathrm{fm}, 1.0~\mathrm{fm}\right] \times \left[0.0~\mathrm{fm}, 1.0~\mathrm{fm}\right]$ with a resolution of 500 cells in each direction and periodic boundary conditions. 
The pressure is uniform with $P = 4/3~\mathrm{GeV}/\mathrm{fm}^3$, while the velocity is initialized with the following form:
  \begin{equation}
    v^x = -\frac{1}{2} \, \sin\!\left(2 \pi y\right) \;,
  \qquad 
    v^y =  \frac{1}{2} \, \sin\!\left(2 \pi x\right) \;.  \label{OrszagTang1}
  \end{equation}
The initial magnetic field is chosen as
  \begin{equation}
    B^x = -B_0 \, \sin\!\left(2 \pi y\right) \;,
  \qquad
    B^y =  B_0 \, \sin\!\left(4 \pi x\right) \;,  \label{OrszagTang2}
  \end{equation}
where $B_0 = 1~\mathrm{GeV}^{1/2}\mathrm{fm}^{-3/2}$. 
The enforcement of the divergence-free constraint is handled with the Flux-Constrained Method (FCT) (see, e.g., Ref.\ \cite{toth:2000b} and references therein). 
The initial profile is extremely unstable and leads to the formation of numerous types of MHD waves, which interact with each other, leading to the complicated vortex pattern that can be seen in Fig.\ \ref{Fig:OrszagTang1}. 
As a comparison, we show in Fig.\ \ref{Fig:OrszagTang1} a simulation in Cartesian coordinates (top row, $t = 1~\mathrm{fm}$, $t_0 = 0~\mathrm{fm}$) and a simulation in Milne coordinates (bottom row, $\tau = 2~\mathrm{fm}$). 
The flow profile is very similar in the two cases, but the pressure and the magnetic field strength in Milne coordinates are weaker than in Minkowski coordinates because the faster decay in Milne coordinates produced by the longitudinal expansion of the system leads to an additional source term in the evolution equations for the energy density and the magnetic field, cf.\ Eq.\ (\ref{BHAC14}). 
In both cases, however, \texttt{BHAC-QGP} is very good at keeping the violation of the divergence constraint to very small values. 
Furthermore, since the initial profile is symmetric under a rotation of $\pi$, Fig.\ \ref{Fig:OrszagTang1} shows that \texttt{BHAC-QGP} can maintain this symmetry in both cases for the newly implemented equation of state.
  \subsection{Ultrarelativistic rotor}
The ``rotor'' problem is a useful test to study the onset and propagation of strong torsional Alfvén waves. 
At the initial time, a cylindrical disk is defined in such a way that it rotates with angular speed $\omega$ relative to a static, external medium. 
This rotation will initiate shear Alfvén waves that propagate along the magnetic field lines. 
These waves are then partially reflected from the surface of the disk back to the origin, where they reflect back and propagate again to the surface of the disk, where they are again reflected and so on. 
This series of reflections slows the rotor down, while it spins up the external medium \cite{Inghirami:2016iru, Stone:2008mh, Mignone:2011fd}. 
\par
We perform this test with \texttt{BHAC-QGP} and the newly implemented ultrarelativistic EOS in Milne coordinates and set the initial time to $\tau_0 = 1.0~\mathrm{fm}$. 
The size of the computational domain is a square with $\left[-0.5~\mathrm{fm}, +0.5~\mathrm{fm}\right] \times \left[-0.5~\mathrm{fm}, +0.5~\mathrm{fm}\right]$ and open boundary conditions on each side. 
We run the test once with a resolution of $200 \times 200$ cells and one level of AMR and once with a resolution of $100 \times 100$ cells but three levels of AMR. 
The rotor itself has a radius of $r_0 = 0.1~\mathrm{fm}$ and the following velocity profile:
  \begin{equation}
    \!\!\!
    \boldsymbol{v} =
      \begin{cases}
        \left(-\omega \dfrac{y}{r_0},~\omega \dfrac{x}{r_0},~0\right) \;, \quad\! &r \le r_0 \;, \\[10pt]
        \left(-f\!\left(r\right) ~\omega \dfrac{y}{r_0}, ~f\!\left(r\right) ~\omega \dfrac{x}{r_0},\, 0\right) \;, \quad\! &r_0 < r \le r_1 \;, \\[14pt]
        \left(0,\, 0,\, 0\right)\;, \quad\! &r_1 < r  \;,
      \end{cases}   \label{Rotor1}
  \end{equation}
where $f := (r_1 - r)/(r_1 - r_0)$ is a damping function with $r_1 = 0.115~\mathrm{fm}$. 
The angular speed is set as $\omega = 0.95$ and the pressure is initialized as
  \begin{equation}
    P = 
      \begin{cases}
        \enspace 10 \;,  \qquad &r \le r_0 \;, \\[4pt]
        \enspace 0.01 \,+\, 9.99 \, f\!\left(r\right) \;,  \qquad &r_0 < r \le r_1 \;, \\[4pt]
        \enspace 0.01 \;,  \qquad &r_1 < r\;.
      \end{cases}    \label{Rotor2}
  \end{equation}
The magnetic field is constant everywhere with $B^x = 5/\sqrt{4 \pi}~\mathrm{GeV}^{1/2}\mathrm{fm}^{-3/2}$ and $B^y = B^z = 0$. 
With these values, the area outside the disc is highly magnetized, and the initial ratio between the magnetic pressure and the fluid pressure, which defines the so-called inverse plasma-$\beta$, is $\beta^{-1} := b^2/(2 P) \approx 100$ outside the disc. 
In such highly magnetized regions, the usual inversion procedures tend to fail, so that \texttt{BHAC-QGP} has to resort to the entropy advection equation. 
Therefore, this setup allows us to test the implemention of the entropy switch. 
\par
\begin{figure*}[!htp]
    \centering
    \includegraphics[width=0.32\linewidth]{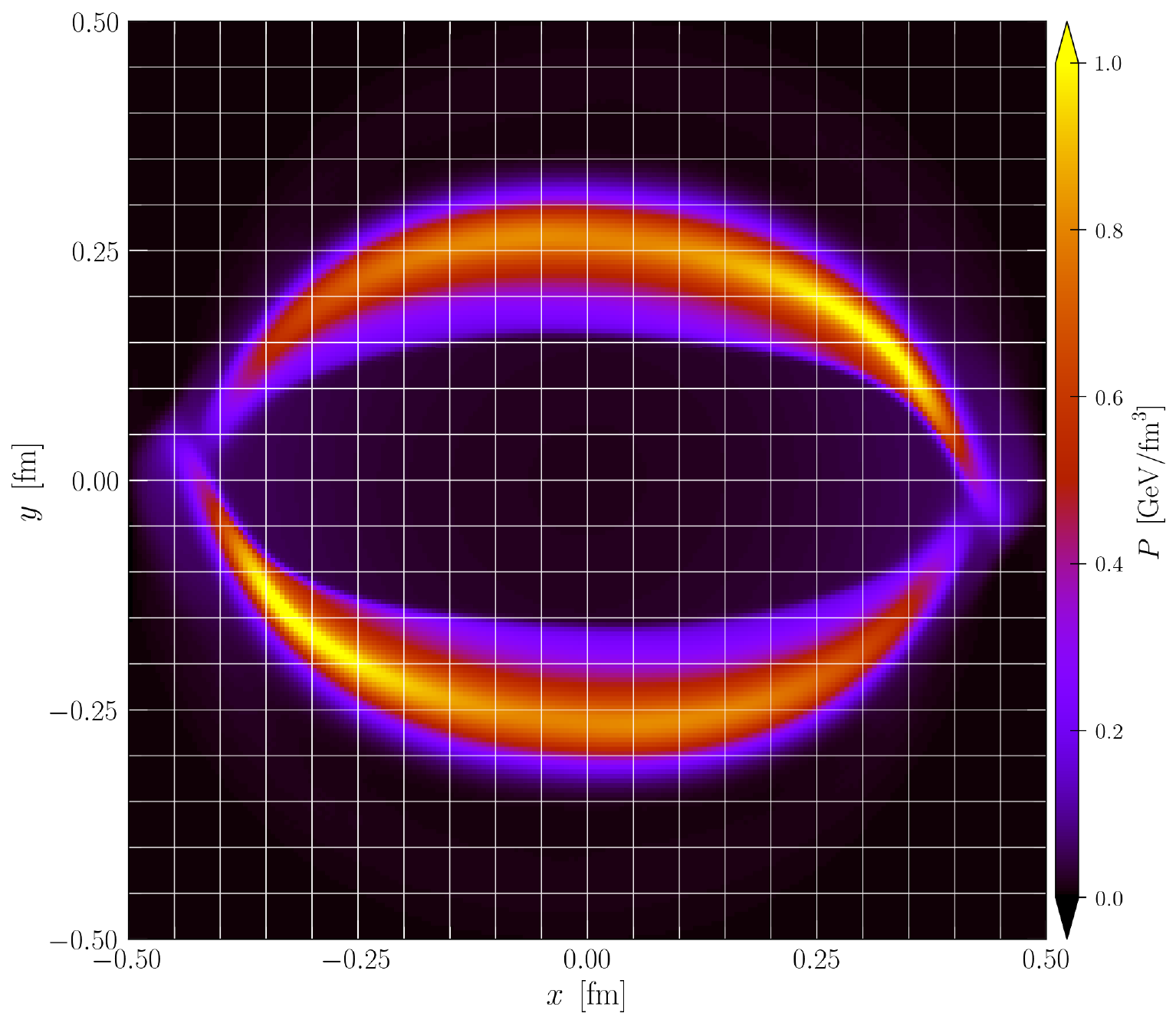}\hfill
    \includegraphics[width=0.32\linewidth]{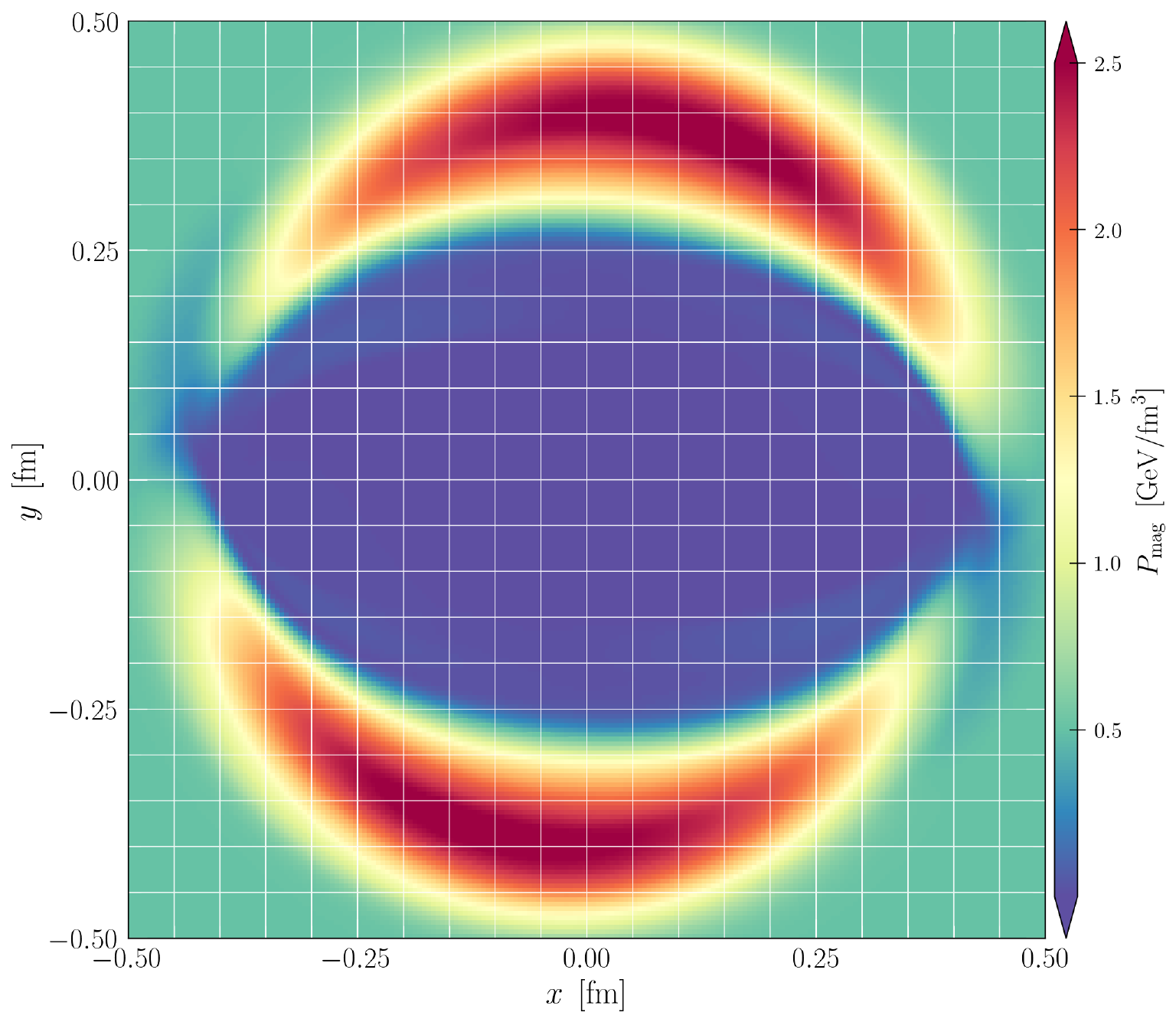}\hfill
    \includegraphics[width=0.32\linewidth]{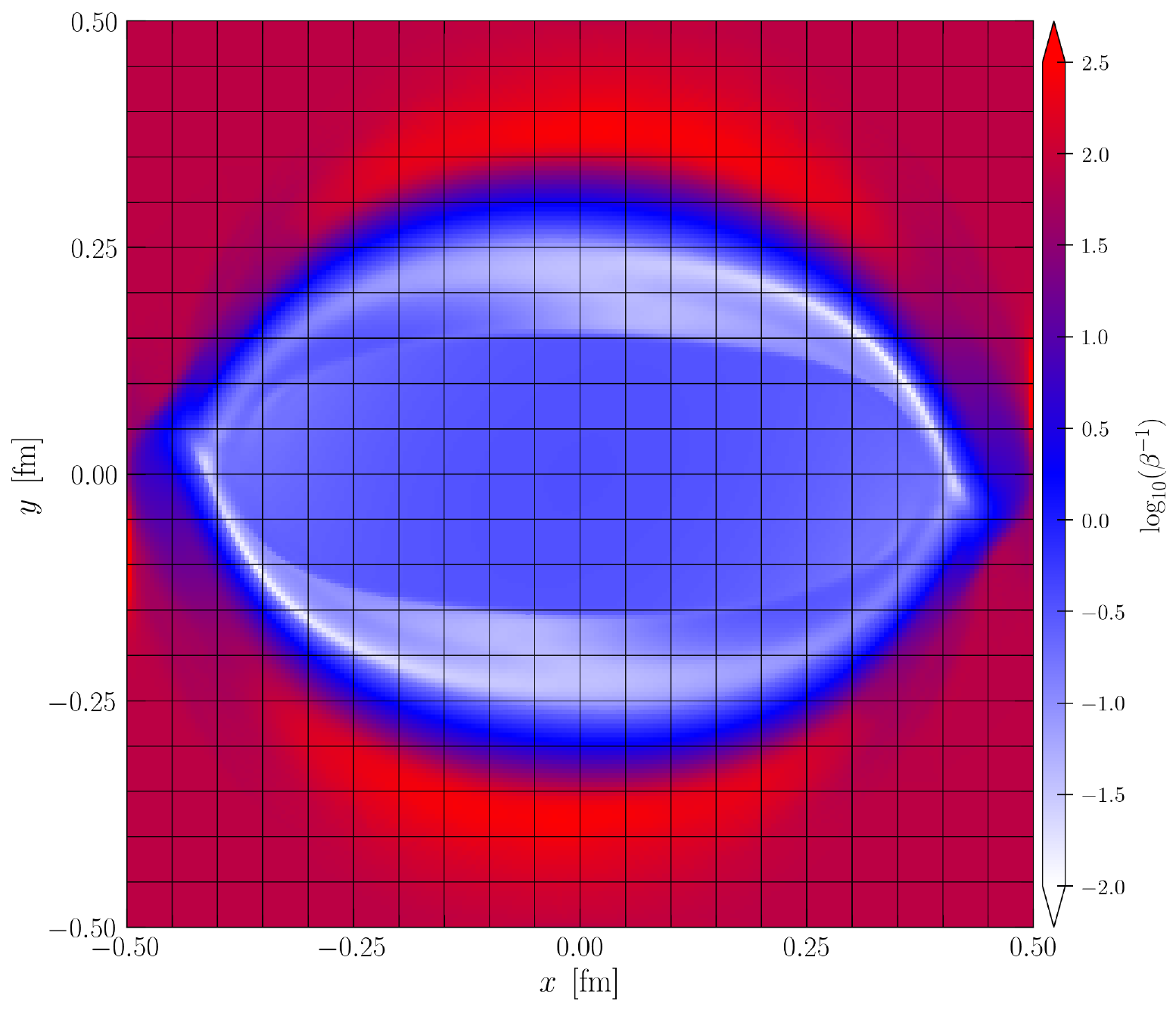} \\
    \includegraphics[width=0.32\linewidth]{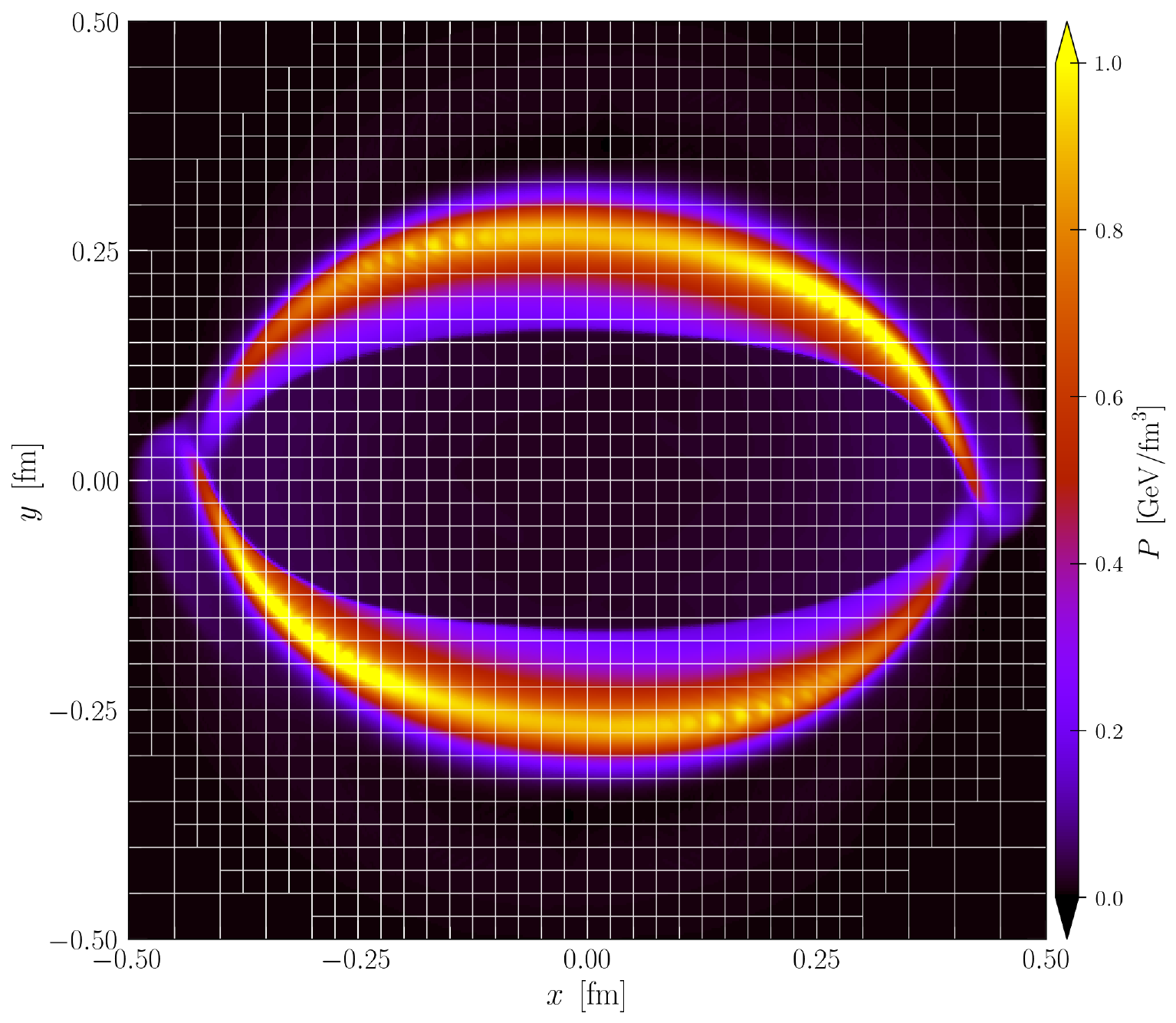}\hfill
    \includegraphics[width=0.32\linewidth]{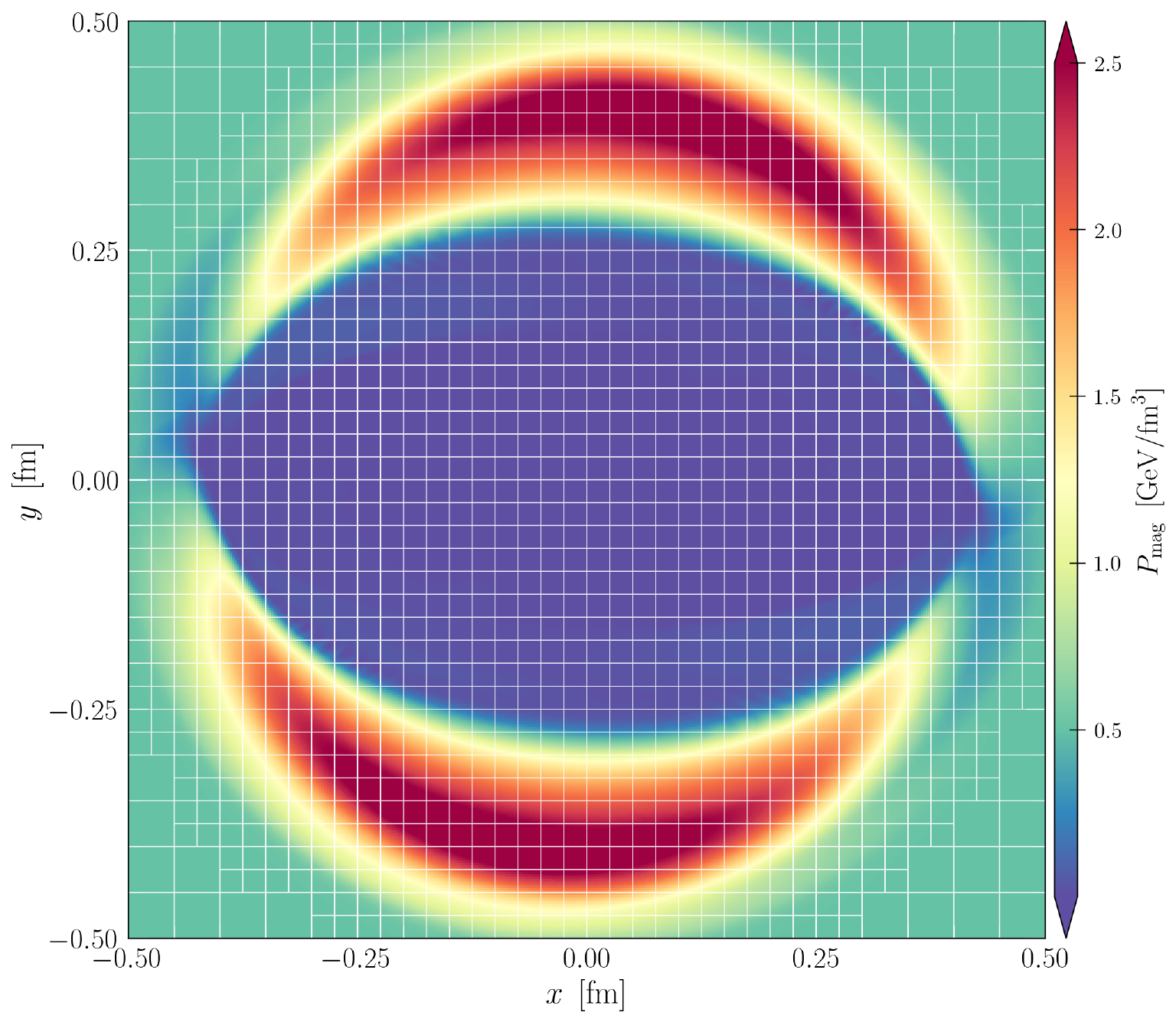}\hfill
    \includegraphics[width=0.32\linewidth]{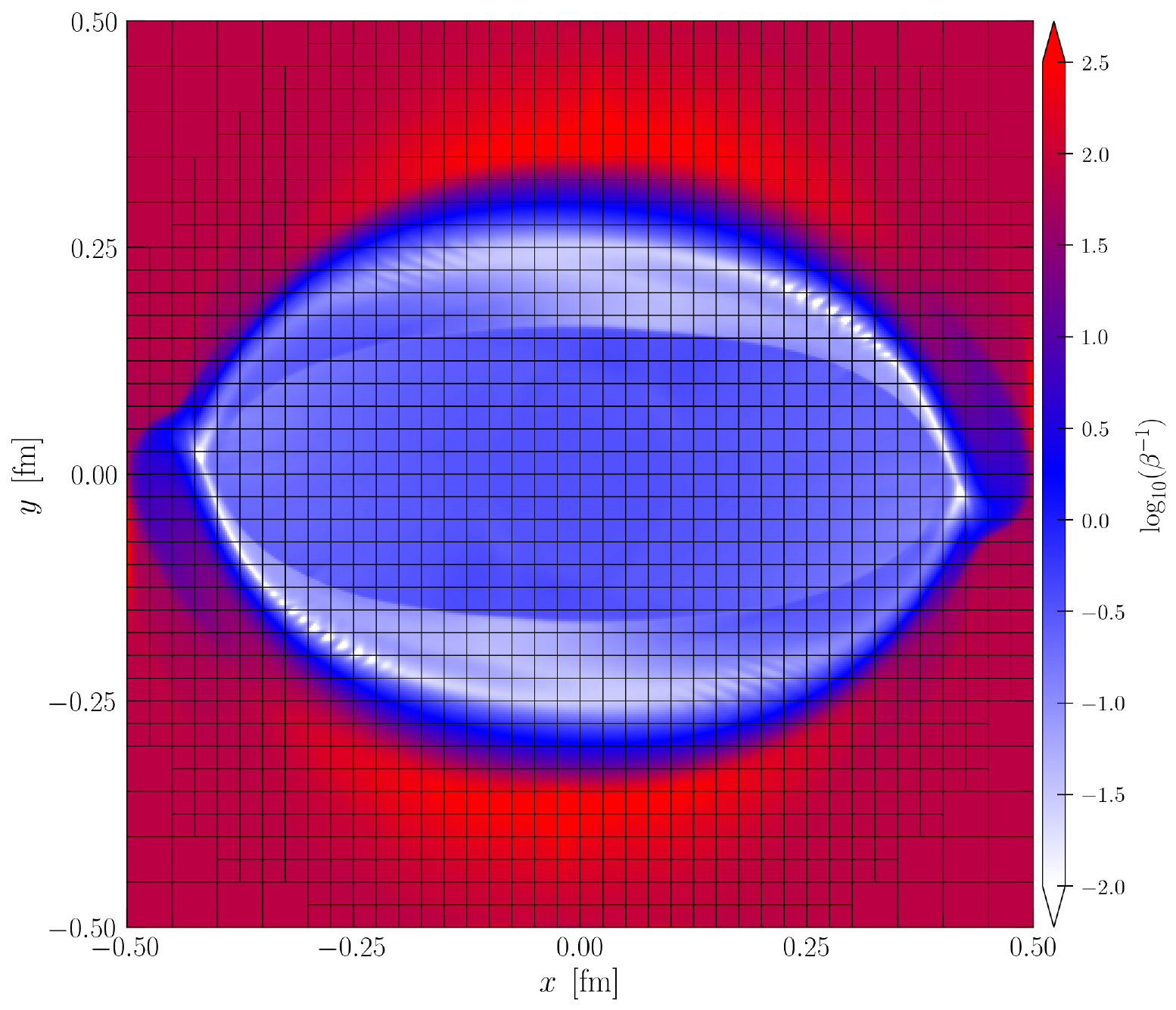}
    \caption{Quantities of the ultrarelativistic rotor problem at $\tau = 1.4~\mathrm{fm}$ (initial time $\tau_0 = 1~\mathrm{fm}$) in Milne coordinates. 
    In the top row the results of a simulation with one level of AMR are plotted, while in the bottom row the corresponding results of a simulation with three levels of AMR are reported. 
    The first column displays the fluid pressure, while the second one shows the magnetic pressure, and the third one the inverse plasma-$\beta$ parameter. 
    With the entropy switch, \texttt{BHAC-QGP} is able to simulate also those regions where $\beta^{-1}$ exceeds the value of 100.}
    \label{Fig:Rotor1}
  \end{figure*}

Figure \ref{Fig:Rotor1} displays quantities of the evolved rotor at $\tau = 1.4~\mathrm{fm}$. 
In the top row we present the results of the simulation with one level of AMR, showing from left to right the fluid pressure, the magnetic pressure, and the inverse beta parameter, while the second row shows the corresponding results with three levels of AMR. 
With the additional AMR levels, individual areas, especially the wavefronts, can be resolved more finely and accurately. 
We note that despite the extreme physical conditions in the test, \texttt{BHAC-QGP} was able to successfully handle this test thanks to the entropy equation (\ref{EntropySwitch1}).
  \subsection{Ultrarelativistic cylindrical blast wave}
A crucial test for all MHD codes is the handling of discontinuities and resulting shock waves, especially in more than one dimension. 
A widely used test in this context is the so-called cylindrical blast-wave problem, where a plasma is expanding into a magnetically dominated region \cite{Stone:2008mh, Mignone:2011fd, DelZanna:2007pk, Komissarov:2007wk}. 
Although there is no exact solution available to compare with, such a test is important because it allows to find weaknesses in a numerical implementation, which might not be seen so clearly in more complicated problems. 
For such a test, a profile is initialized in such a way that a strong shock wave is moving outwards as the system evolves. 
Shock waves may lead to unphysical values in quantities like the pressure or the magnitude of the velocity. 
Therefore, a very robust shock-capturing technique is required in order to obtain a stable simulation. 
\par
For our setup we define a computational grid of $\left[-0.5~\mathrm{fm}, +0.5~\mathrm{fm}\right] \times \left[-0.5~\mathrm{fm}, +0.5~\mathrm{fm}\right]$.
We perform this test once with one level of AMR and a resolution of $200 \times 200$ cells and once with three levels of AMR but a resolution of $100 \times 100$ cells. 
The whole domain contains a uniform magnetic field 
$B^x = B^y = \sqrt{2}~\mathrm{GeV}^{1/2}\mathrm{fm}^{-3/2}$ and an initial velocity profile that is zero everywhere. 
At the center of the grid we initialize a cylinder of radius $r = \sqrt{x^2 + y^2} = 0.1~\mathrm{fm}$ where $P = 10~\mathrm{GeV}/\mathrm{fm}^3$. 
The pressure in the ambient medium is set to be $P = 0.01~\mathrm{GeV}/\mathrm{fm}^3$. 
Hence, the region outside the cylinder is highly magnetic, with $\beta^{-1} = P_{\mathrm{mag}}/P = 200$. 
Usual inversion schemes are likely to fail in this region, so that \texttt{BHAC-QGP} has to rely on the entropy switch. 
The initial time is set to be $\tau_0 = 1~\mathrm{fm}$ and the divergence-free constraint is maintained by the Flux Constraint Transport (FCT).
\par
\begin{figure*}[!htp]
    \centering
    \includegraphics[width=0.32\linewidth]{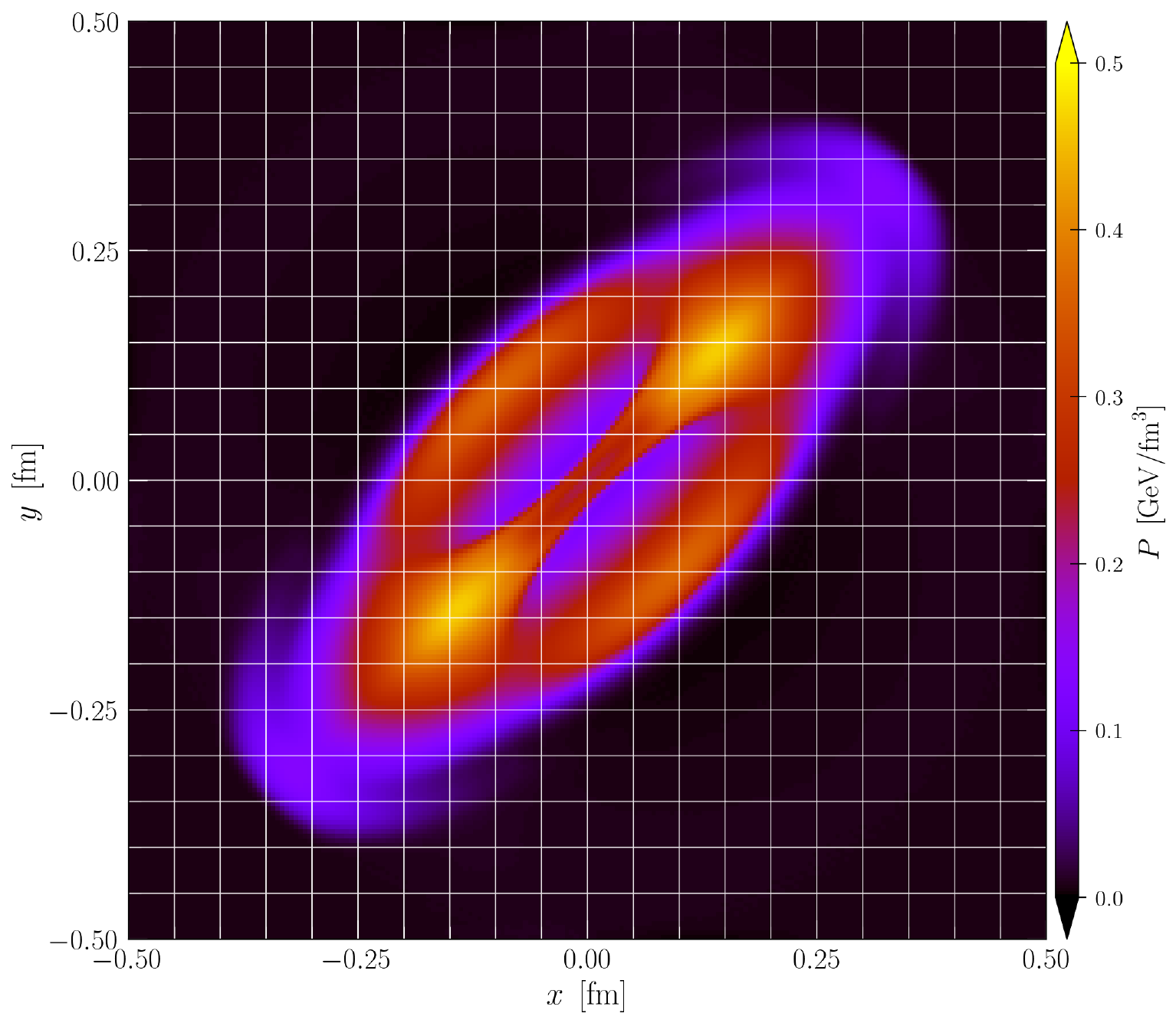}\hfill
    \includegraphics[width=0.32\linewidth]{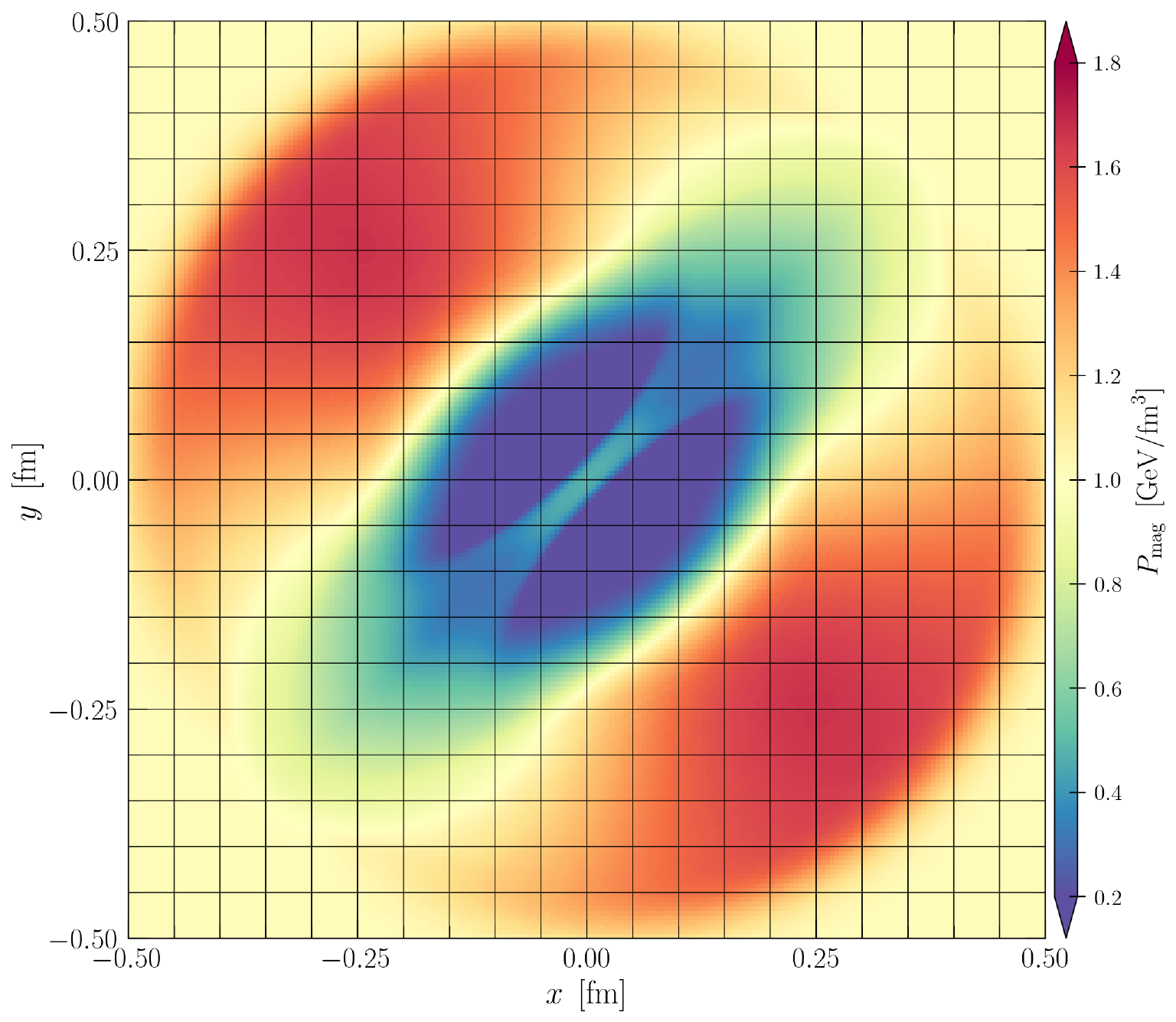}\hfill
    \includegraphics[width=0.32\linewidth]{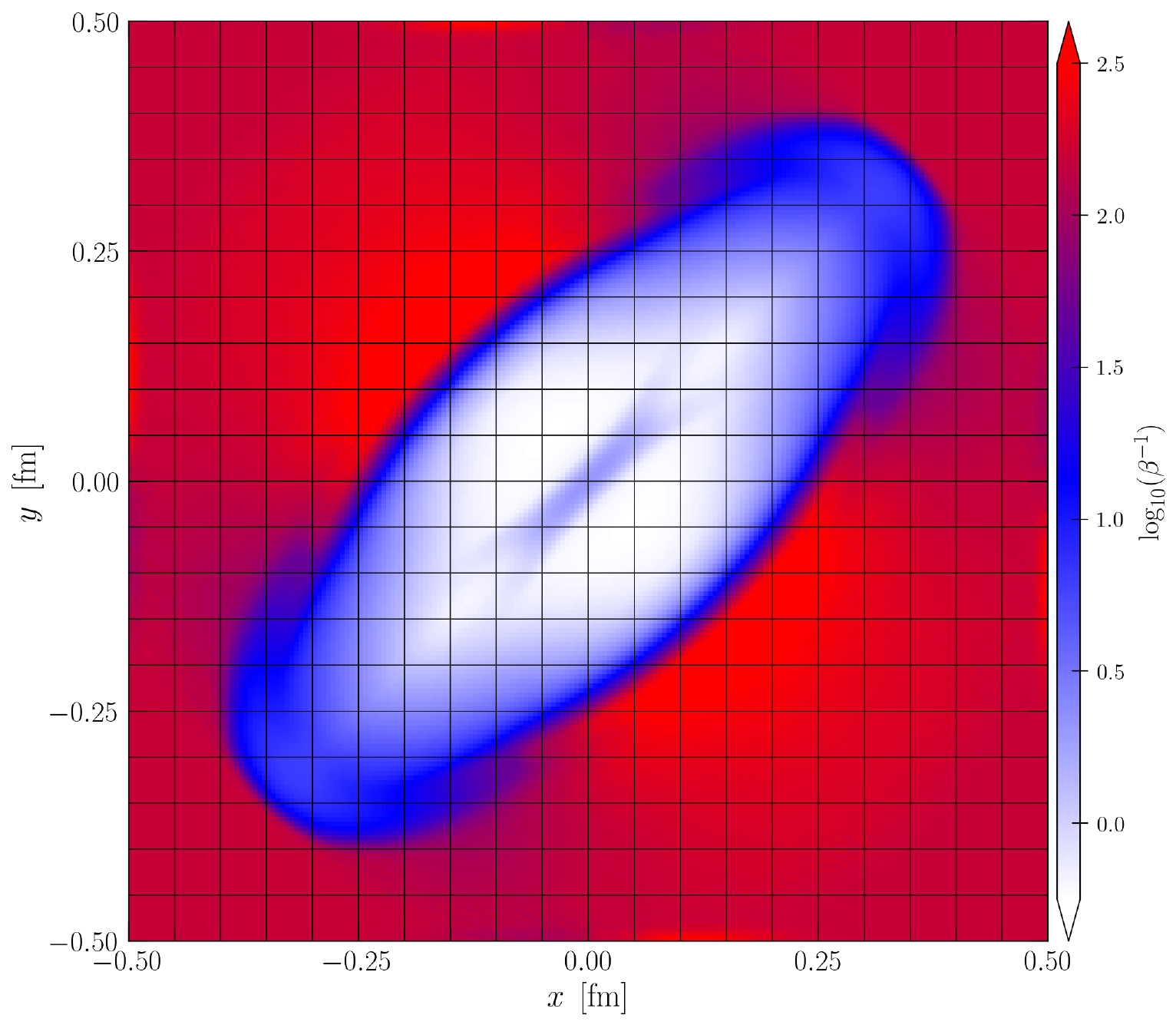} \\
    \includegraphics[width=0.32\linewidth]{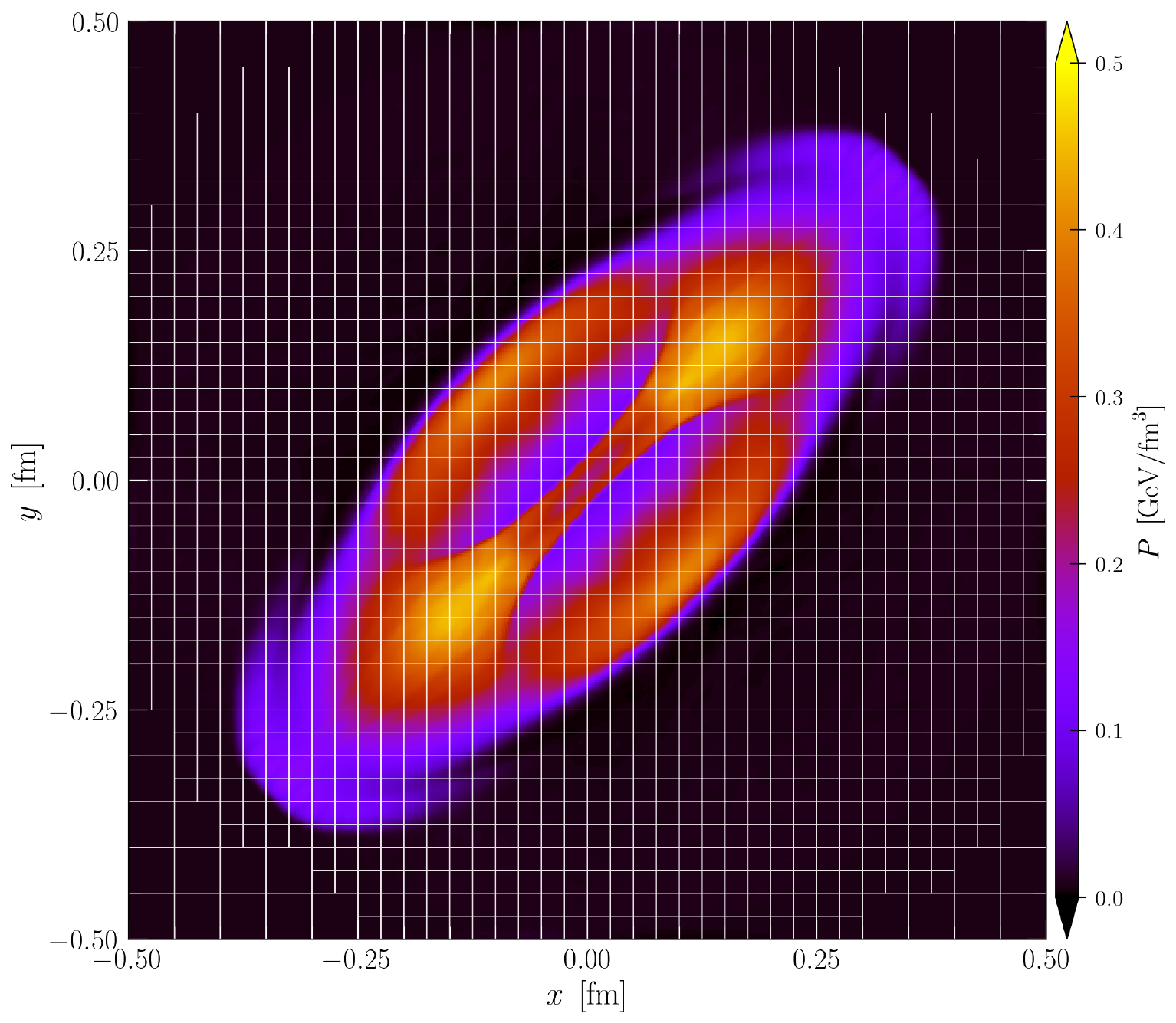}\hfill
    \includegraphics[width=0.32\linewidth]{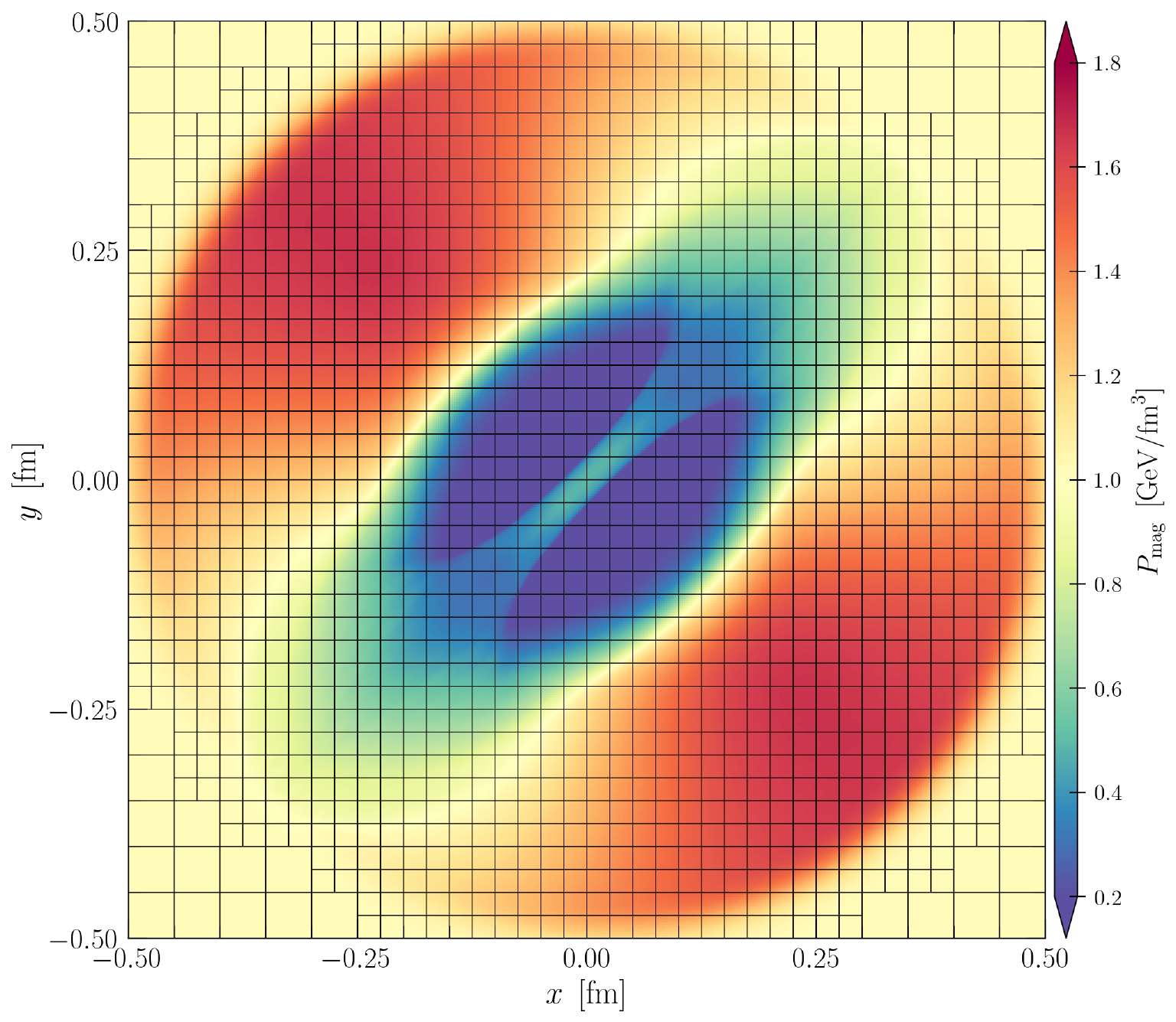}\hfill
    \includegraphics[width=0.32\linewidth]{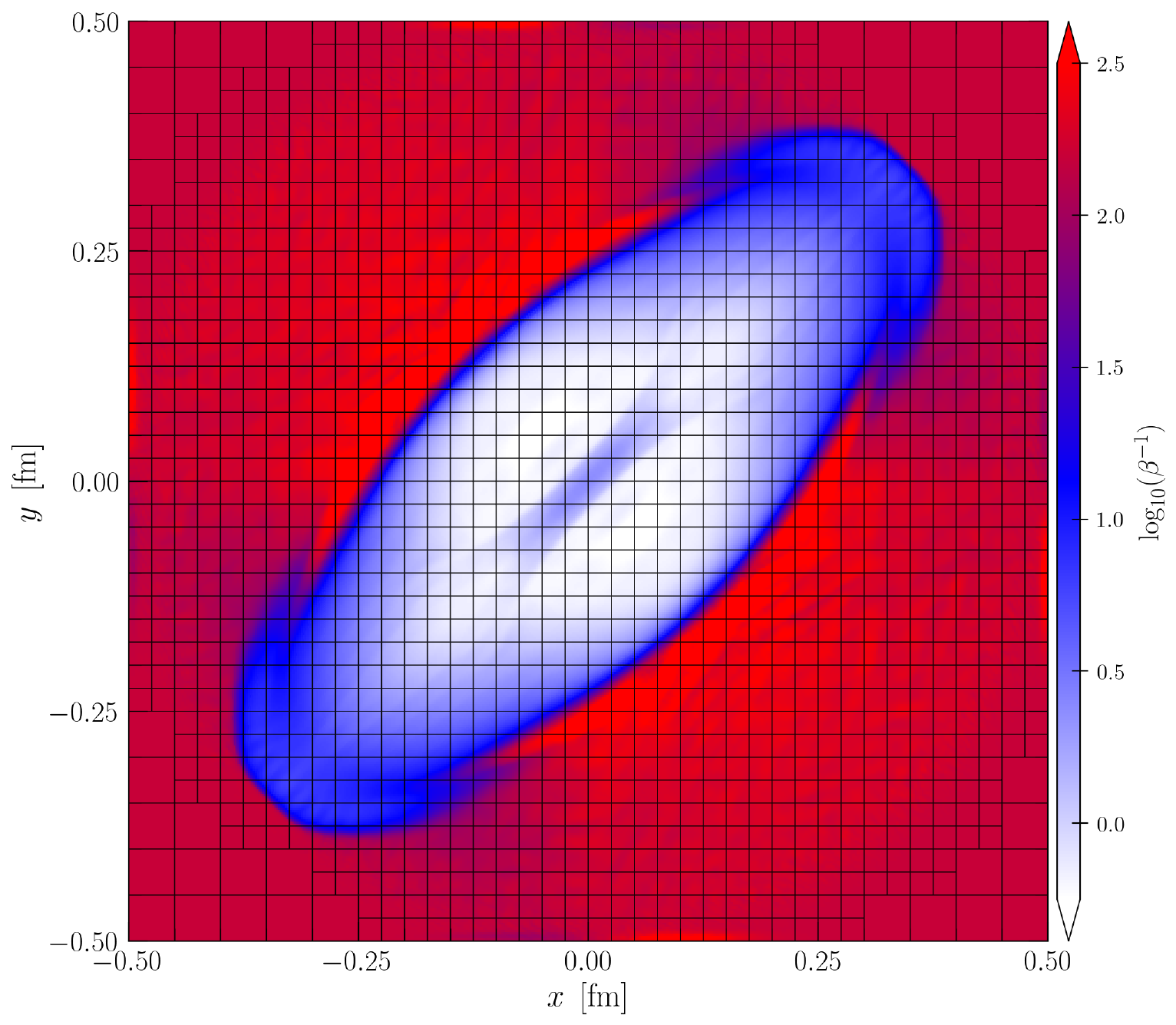}
    \caption{Results for the cylindrical blast-wave test at $\tau = 1.4~\mathrm{fm}$ (initial time $\tau_0 = 1~\mathrm{fm}$) in Milne coordinates. 
    In the first row the results of a simulation with one level of AMR are plotted, while in the bottom row the corresponding results of a simulation with three levels of AMR are reported. 
    The first column displays the pressure, while the second one shows the magnetic pressure, and the third one the inverse plasma-$\beta$ parameter. 
    With the entropy switch, \texttt{BHAC-QGP} is able to simulate also those regions where $\beta^{-1}$ exceeds the value of 100.}
    \label{Fig:CylindricalExplosion1}
  \end{figure*}

Figure \ref{Fig:CylindricalExplosion1} displays the results of this test at $\tau = 1.4~\mathrm{fm}$, with the panels showing one fast outgoing shock wave which is circular and a rarefaction wave that is traveling inwards. 
Also in this case, strong shocks and regions with inverse $\beta$-parameter in excess of $100$ are handled without problems in Milne coordinates.
  \subsection{Ultrarelativistic spherical explosion}
As a final test driven by our ultimate goal of simulating heavy-ion collisions with \texttt{BHAC-QGP}, we present the three-dimensional spherical-explosion problem \cite{Stone:2008mh, Mignone:2011fd, Komissarov:2007wk}. 
Although heavy-ion collisions do not produce shock waves as strong as in astrophysical phenomena, the region outside the fireball is highly magnetized due to the strong external electromagnetic fields. 
Consequently, \texttt{BHAC-QGP} must be able to run simulations in Milne coordinates, in which there are areas where the magnetic pressure is far stronger than the pressure of the fluid. 
Additionally, such a multidimensional problem allows us to investigate the performance of \texttt{BHAC-QGP} in a fully three-dimensional context. 
Just like the equivalent two-dimensional cylindrical blast-wave test, the spherical explosion must prove that the code can handle the formation and propagation of all types of MHD waves. 
Especially when the numerical scheme cannot control the divergence-free constraint sufficiently, unphysical states could be obtained involving negative pressure because the background magnetic pressure increases the strength of magnetic monopoles. 
In the multidimensional relativistic case, it is very difficult to treat situations where the Alfvén velocities are close to the speed of light, because numerical errors act independently on velocity and magnetic-field components. 
Therefore, this can easily lead to incorrect fluxes and eventually yields unphysical states, for example that $v^2 \ge 1$ when primitive variables are recovered from the evolved conservative ones. 
Moreover, terms in the total energy equation are strongly unbalanced in these cases, and, again, numerical errors may lead to code crashing. 
This happens especially in highly magnetic regions, where the magnetic pressure is at least $100$ times higher than the kinetic pressure. 
In such cases, it is useful to discard the energy equation and use the entropy advection equation instead. 
\par
For this test we define a computational domain of 
$\left[-0.5~\mathrm{fm}, +0.5~\mathrm{fm}\right] \times \left[-0.5~\mathrm{fm}, +0.5~\mathrm{fm}\right] \times \left[-0.5~\mathrm{fm}, +0.5~\mathrm{fm}\right]$. 
The numerical scheme is ``minmod'' and RK3 for time stepping. 
The center of the domain contains an over-pressurized, $P = 10~\mathrm{GeV}/\mathrm{fm}^3$, spherical region with radius $r = 0.1~\mathrm{fm}$, while the pressure in the exterior region is set to $P = 0.01~\mathrm{GeV}/\mathrm{fm}^3$. 
The strength of the uniform magnetic field is set to $B^x = B^y = B^z = 1~\mathrm{GeV}^{1/2}\mathrm{fm}^{-3/2}$. 
This puts the value of the inverse plasma-$\beta$ to $150$ outside the sphere, making it a challenging test for \texttt{BHAC-QGP}.
The initial velocity is set to zero everywhere.
\par
\begin{figure*}
    \centering
    \includegraphics[width=0.47\textwidth]{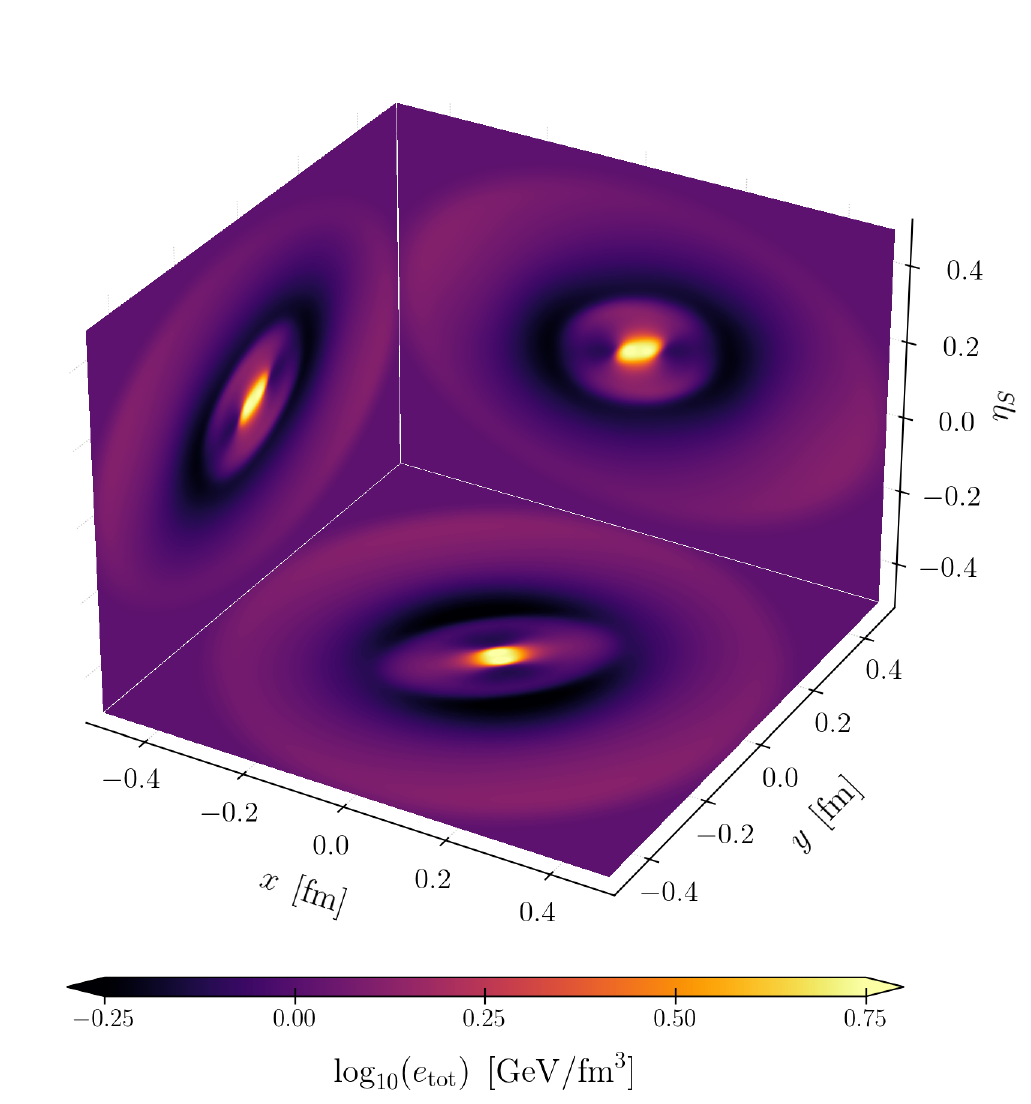}\hfill
    \includegraphics[width=0.47\textwidth]{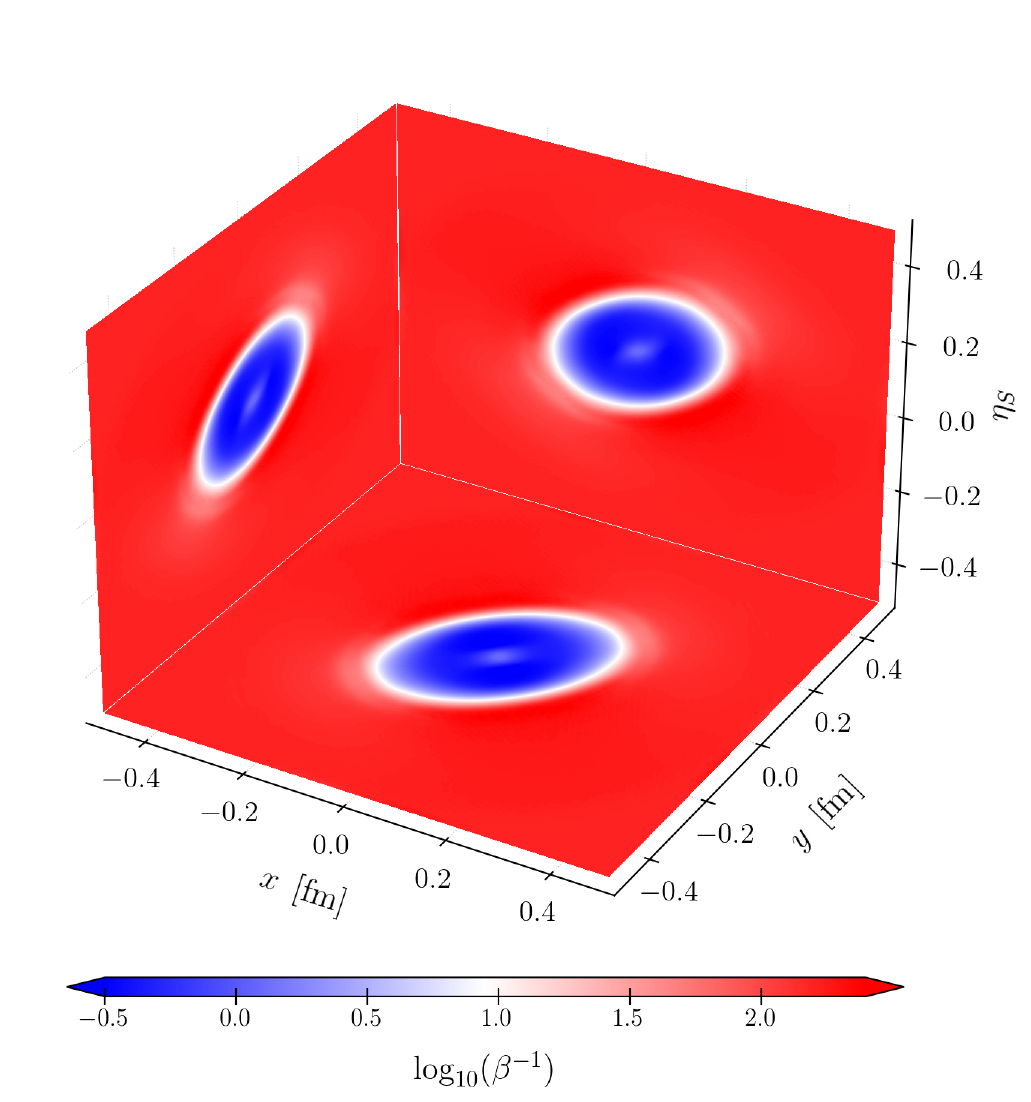}
    \caption{Total energy density (left panel) and inverse plasma-$\beta$ (right panel) of the spherical explosion test at $\tau = 1.4~\mathrm{fm}$ (initial time $\tau_0 = 1~\mathrm{fm}$). 
    The second plot shows that \texttt{BHAC-QGP} can also simulate in \textcolor{blue}{three dimensions \sout{3D the}} regions where the magnetic pressure exceeds that of the fluid by several orders of magnitude. 
    The projections in the $x$-$y$ plane are taken at $\eta_S = 0~\mathrm{fm}$ while those for the $y$-$\eta_S$ plane are obtained at $x = 0~\mathrm{fm}$. 
    In the same way, the plots in the $x$-$\eta_S$ plane are calculated at $y = 0~\mathrm{fm}$.}
    \label{Fig:SphericalExplosion1}
  \end{figure*}

Results of this test at $\tau = 1.4~\mathrm{fm}$ are given in Fig.\ \ref{Fig:SphericalExplosion1}. 
The general structure of the solution is similar to that of the cylindrical case, but the central rarefaction wave is much more pronounced. 
Clearly and satisfactorily, \texttt{BHAC-QGP} is able to calculate even the highly magnetized regions. 
Due to the longitudinal expansion, the circular symmetry is lost along the $\eta_S$-direction, while the expansion in the $x$-$y$-plane is still symmetric.
\section{Conclusion \& Outlook}
\label{ConclusionOutlook}
We introduced \texttt{BHAC-QGP}, a novel code derived from the astrophysical code \texttt{BHAC}, that is now also able to solve the equations of ideal GRMHD in three spatial dimensions in Milne coordinates and flat spacetimes. 
Since the main motivation of \texttt{BHAC-QGP} is to simulate the spacetime evolution of matter produced in heavy-ion collisions, we also implemented an equation of state for massless particles. 
\texttt{BHAC-QGP} shares many features with the original code, especially Adaptive Mesh Refinement, allowing us to dynamically adjust the computational resolution whenever necessary, which consequently leads to particularly efficient calculations. 
\texttt{BHAC-QGP} has undergone a number of serious tests that have been presented in this paper, and where the numerical results are compared to the analytic ones whenever possible.
In all cases, the tests have been passed successfully and accurately. 
Furthermore, we have carried out multidimensional tests, which especially show \texttt{BHAC-QGP}'s ability to handle also highly magnetic areas with the help of the entropy switch. 
This will enable us to simulate the dynamical evolution of the medium formed in heavy-ion collisions, consistently taking into account the interplay with the strong magnetic fields generated in such an environment.
\section*{Acknowledgments}
The authors acknowledge support by the Deutsche Forschungsgemeinschaft (DFG, German Research Foundation) through the CRC-TR 211 ``Strong-interaction matter under extreme conditions'' – project number 315477589 – TRR 211.
The work is supported by the State of Hesse within the Research Cluster ELEMENTS (Project ID 500/10.006). 
Computational resources have been provided by the Center for Scientific Computing (CSC) at the Goethe University.
M.M.\ would like to thank H.\ Olivares, M.\ Chabanov, N.\ Kuebler, and J.\ Sammet for fruitful discussions.
\bibliography{biblio1}  


\end{document}